\documentclass[preprint,aps,showpacs,showkeys,nofootinbib]{revtex4}
\usepackage{graphicx}
\usepackage{dcolumn}
\usepackage{bm}
\begin{document}
\title{Lepton flavor violation in the BLMSSM}
\author{Shu-Min Zhao$^{1}$\footnote{zhaosm@hbu.edu.cn}, Tai-Fu Feng$^{1,2}$\footnote{fengtf@hbu.edu.cn}, Hai-Bin Zhang$^{1}$,  Xi-Jie Zhan$^{1}$, Yin-Jie Zhang$^1$, Ben Yan$^1$}
\affiliation{$^1$ Department of Physics, Hebei University, Baoding
071002,
China\\
$^2$ State Key Laboratory of Theoretical Physics, Institute of Theoretical Physics, Chinese Academy of Sciences,
Beijing 100190, China}
\date{\today}
\begin{abstract}
In a supersymmetric extension
of the standard model with local gauged baryon
and lepton numbers (BLMSSM), there are new sources for lepton flavor violation, because the right-handed neutrinos and new gauginos are
introduced. In BLMSSM, we
study the charged lepton flavor violating processes $l_j\rightarrow l_i+\gamma$ and $l_j\rightarrow 3l_i$ in detail.
The numerical results show that in some parameter space the branching ratios for charged lepton flavor violating processes can be large enough
to be detected in the near future.
\end{abstract}

\maketitle
\section{Introduction}
From the neutrino oscillation experiments\cite{neutrino}, it is convincing that neutrinos have tiny masses and mix with each
other\cite{neutrinoN1,neutrinoN2}.
Therefore, lepton flavor symmetry is not exact in the universe. Though the standard model(SM) has achieved great success with the detected
lightest CP-even Higgs, SM should be extended.
Because of the GIM mechanism, in SM the charged lepton flavor violating(CFLV) processes are very tiny, for example $Br_{SM}(l_j\to l_i+\gamma)\sim 10^{-55}$\cite{mueg}.
The experiment upper bounds of the CLFV processes $l_j\to l_i+\gamma$ and $l_j\to 3l_i $ are\cite{14pdg}
\begin{eqnarray}
&&\texttt{Br}(\mu\rightarrow e\gamma)<5.7\times10^{-13},~~~~~\texttt{Br}(\mu\rightarrow 3e)<1.0\times10^{-12},
\nonumber\\&&\texttt{Br}(\tau\rightarrow e\gamma)<3.3\times10^{-8},~~~~~\texttt{Br}(\tau\rightarrow \mu\gamma)<4.4\times10^{-8},
\nonumber\\&&\texttt{Br}(\tau\rightarrow 3e)<2.7\times10^{-8},~~~~~\texttt{Br}(\tau\rightarrow 3\mu)<2.1\times10^{-8}.
\end{eqnarray}
They are much larger than the corresponding SM theoretical predictions. To explore new physics beyond SM, study CLFV processes is an effective
approach.
Once physicists observe CLFV processes in future experiments, there must be new physics beyond SM.

In a simple extension of SM, with a new additional Yukawa matrix for  right-handed neutrinos, CLFV processes are induced
at loop level with neutrino\cite{RNeutrino}. They are suppressed strong by the tiny neutrino masses and impossible to be observed practically.
One popular supersymmetric extension of SM is the minimal suppersymmetric standard model(MSSM)\cite{MSSM}. In R-parity conserved MSSM, the
left-handed light neutrinos are still massless and can not explain the discovery of neutrino oscillations. Therefore, physicists
extend MSSM to account for the light neutrino masses and mixings. Adding low-scale right-handed neutrinos and approximate lepton
number symmetries, $\nu_R$MSSM is obtained, where the authors study the CLFV processes\cite{nuRMSSM}.  In the supersymmetric standard
model with right-handed neutrino supermultiplets, the authors investigate various LFV processes in detail\cite{ljlig}.
In our previous work, we study neutrino masses and CLFV processes in $\mu\nu$SSM\cite{ZHB}.

For the beyond SM models, one can violate R parity\cite{Rp} with the non-conservation of baryon number ($B$) or lepton number ($L$)\cite{BLMSSM,Rp1}.
A minimal supersymmetric extension of
the SM with local gauged $B$ and $L$(BLMSSM) is a favorite one\cite{BLMSSM1}. BLMSSM was first proposed in one
of the references in \cite{BLMSSM1}. In the work, this model is that we are adopting.
The local gauged $B$ is used to explain the matter-antimatter asymmetry in the universe.
Right-handed neutrinos are introduced in BLMSSM to account for the neutrino oscillation experiments, which lead to three tiny neutrino masses
through the seesaw mechanism.
Then lepton number ($L$) is expected to be broken spontaneously around ${\rm TeV}$ scale.
In BLMSSM, the lightest CP-even Higgs mass and the decays $h^0\rightarrow\gamma\gamma$, $h^0\rightarrow ZZ (WW)$ are studied
in the work\cite{weBLMSSM}. Taking into account the Yukawa couplings between Higgs and exotic quarks, we study the
neutron and lepton electric dipole moments(EDMs) in the CP-violating BLMSSM\cite{smneutron,BLLEDM}.
 $B^0-\bar{B}^0$ mixing and $t\rightarrow c+\gamma,t\rightarrow c+g$ are
also investigated in SM extension with local gauged baryon and lepton numbers\cite{sunfei}.

In this work, we analyze these CLFV processes
($\mu\rightarrow e\gamma$, $\mu\rightarrow3e$; $\tau\rightarrow e\gamma$, $\tau\rightarrow\mu\gamma$,
$\tau\rightarrow3e$,$\tau\rightarrow3\mu$) in the frame work of BLMSSM. Compared with MSSM, there
are new sources to enlarge these CLFV processes via loop contributions. The new CLFV scores are produced from
1. the right-handed neutrinos mixing with left-handed neutrinos;
2. the coupling of new neutralino(lepton neutralino)-slepton-lepton.
In some parameter space of BLMSSM, large corrections to the CLFV processes are obtained, and can easily exceed their experiment upper bounds.
Therefore, to enhance these CLFV processes is possible, and they may be measured in the near future.

After this introduction, we briefly summarize the main ingredients
of the BLMSSM, and show the needed mass matrices and couplings in section 2.
In section 3, the decay widths of these interested CLFV processes are analyzed.
The input parameters and numerical analysis are shown in section 4 and we give our conclusion in section 5.

\section{BLMSSM}
BLMSSM is the supersymmetric extension of the SM with local gauged $B$ and $L$, whose local gauge group is $SU(3)_{C}\otimes SU(2)_{L}\otimes U(1)_{Y}\otimes U(1)_{B}\otimes U(1)_{L}$\cite{BLMSSM}.
The exotic leptons $\hat{L}_{4}\sim(1,\;2,\;-1/2,\;0,\;L_{4})$,
$\hat{E}_{4}^c\sim(1,\;1,\;1,\;0,\;-L_{4})$, $\hat{N}_{4}^c\sim(1,\;1,\;0,\;0,\;-L_{4})$,
$\hat{L}_{5}^c\sim(1,\;2,\;1/2,\;0,\;-(3+L_{4}))$, $\hat{E}_{5}\sim(1,\;1,\;-1,\;0,\;3+L_{4})$ and
$\hat{N}_{5}\sim(1,\;1,\;0,\;0,\;3+L_{4})$ are introduced  to cancel $L$ anomaly. As well as,
the exotic quarks $\hat{Q}_{4}\sim(3,\;2,\;1/6,\;B_{4},\;0)$,
$\hat{U}_{4}^c\sim(\bar{3},\;1,\;-2/3,\;-B_{4},\;0)$,
$\hat{D}_{4}^c\sim(\bar{3},\;1,\;1/3,\;-B_{4},\;0)$,
$\hat{Q}_{5}^c\sim(\bar{3},\;2,\;-1/6,\;-(1+B_{4}),\;0)$, $\hat{U}_{5}\sim(3,\;1,\;2/3,\;1+B_{4},\;0)$ and
$\hat{D}_{5}\sim(3,\;1,\;-1/3,\;1+B_{4},\;0)$ are introduced  to cancel $B$ anomaly.
To break lepton number and baryon number spontaneously, the Higgs superfields $\hat{\Phi}_{L}(1, 1, 0, 0, -2),\hat{\varphi}_{L}(1, 1, 0, 0, 2)$  and
$\hat{\Phi}_{B}(1, 1, 0, 1, 0),\hat{\varphi}_{B}(1, 1, 0, -1, 0)$ are introduced respectively. Now, Higgs mechanism is the very massy foundtion stone for particle physics, and
people are convinced of it, because of the detection of the lightest CP even Higgs $h^0$ at LHC\cite{Higgs}.
The Higgs fields $\hat{\Phi}_{L},\hat{\varphi}_{L}$  and
$\hat{\Phi}_{B},\hat{\varphi}_{B}$ acquire nonzero vacuum expectation values (VEVs), then exotic leptons and exotic quarks obtain masses.
In the BLMSSM, the superfields $\hat{X}(1, 1, 0, 2/3+B_4, 0)$,
$\hat{X}^\prime(1, 1, 0,-(2/3+B_4), 0)$ are introduced to make the heavy exotic quarks unstable.
Furthermore $\hat{X}$ and $\hat{X}^\prime$ mix together, where the lightest mass eigenstate can be a candidate for dark matter.

 The superpotential of BLMSSM is\cite{weBLMSSM}
\begin{eqnarray}
&&{\cal W}_{{BLMSSM}}={\cal W}_{{MSSM}}+{\cal W}_{B}+{\cal W}_{L}+{\cal W}_{X}\;,
\label{superpotential1}
%%%%%%%%%%%%%%%%%%%%%%%%%%%%%%%%%%%%%%%
%%%%%%%%%%%%%%%%%%%%%%%%%%%%%%%%%%%%%%%
\nonumber\\&&{\cal W}_{B}=\lambda_{Q}\hat{Q}_{4}\hat{Q}_{5}^c\hat{\Phi}_{B}+\lambda_{U}\hat{U}_{4}^c\hat{U}_{5}
\hat{\varphi}_{B}+\lambda_{D}\hat{D}_{4}^c\hat{D}_{5}\hat{\varphi}_{B}+\mu_{B}\hat{\Phi}_{B}\hat{\varphi}_{B}
\nonumber\\
&&\hspace{1.2cm}
+Y_{{u_4}}\hat{Q}_{4}\hat{H}_{u}\hat{U}_{4}^c+Y_{{d_4}}\hat{Q}_{4}\hat{H}_{d}\hat{D}_{4}^c
+Y_{{u_5}}\hat{Q}_{5}^c\hat{H}_{d}\hat{U}_{5}+Y_{{d_5}}\hat{Q}_{5}^c\hat{H}_{u}\hat{D}_{5}\;,
\nonumber\\
%%%%%%%%%%%%%%%%%%%%%%%%%%%%%%%%%%%%%%%%%%%%%%%%%%%%%%%%%%%%%
&&{\cal W}_{L}=Y_{{e_4}}\hat{L}_{4}\hat{H}_{d}\hat{E}_{4}^c+Y_{{\nu_4}}\hat{L}_{4}\hat{H}_{u}\hat{N}_{4}^c
+Y_{{e_5}}\hat{L}_{5}^c\hat{H}_{u}\hat{E}_{5}+Y_{{\nu_5}}\hat{L}_{5}^c\hat{H}_{d}\hat{N}_{5}
\nonumber\\
&&\hspace{1.2cm}
+Y_{\nu}\hat{L}\hat{H}_{u}\hat{N}^c+\lambda_{{N^c}}\hat{N}^c\hat{N}^c\hat{\varphi}_{L}
+\mu_{L}\hat{\Phi}_{L}\hat{\varphi}_{L}\;,
\nonumber\\
%%%%%%%%%%%%%%%%%%%%%%%%%%%%%%%%%%%%%%%%%%%%%%%%%%%%%%%%%%%%%
&&{\cal W}_{X}=\lambda_1\hat{Q}\hat{Q}_{5}^c\hat{X}+\lambda_2\hat{U}^c\hat{U}_{5}\hat{X}^\prime
+\lambda_3\hat{D}^c\hat{D}_{5}\hat{X}^\prime+\mu_{X}\hat{X}\hat{X}^\prime\;.
\label{superpotential-BL}
\end{eqnarray}
where ${\cal W}_{{MSSM}}$ is the superpotential of the MSSM.
The soft breaking terms $\mathcal{L}_{{soft}}$ of the BLMSSM can be found in the works\cite{BLMSSM1, weBLMSSM}
\begin{eqnarray}
&&{\cal L}_{{soft}}={\cal L}_{{soft}}^{MSSM}-(m_{{\tilde{N}^c}}^2)_{{IJ}}\tilde{N}_I^{c*}\tilde{N}_J^c
-m_{{\tilde{Q}_4}}^2\tilde{Q}_{4}^\dagger\tilde{Q}_{4}-m_{{\tilde{U}_4}}^2\tilde{U}_{4}^{c*}\tilde{U}_{4}^c
-m_{{\tilde{D}_4}}^2\tilde{D}_{4}^{c*}\tilde{D}_{4}^c
\nonumber\\
&&\hspace{1.3cm}
-m_{{\tilde{Q}_5}}^2\tilde{Q}_{5}^{c\dagger}\tilde{Q}_{5}^c-m_{{\tilde{U}_5}}^2\tilde{U}_{5}^*\tilde{U}_{5}
-m_{{\tilde{D}_5}}^2\tilde{D}_{5}^*\tilde{D}_{5}-m_{{\tilde{L}_4}}^2\tilde{L}_{4}^\dagger\tilde{L}_{4}
-m_{{\tilde{\nu}_4}}^2\tilde{N}_{4}^{c*}\tilde{N}_{4}^c
\nonumber\\
&&\hspace{1.3cm}
-m_{{\tilde{e}_4}}^2\tilde{E}_{_4}^{c*}\tilde{E}_{4}^c-m_{{\tilde{L}_5}}^2\tilde{L}_{5}^{c\dagger}\tilde{L}_{5}^c
-m_{{\tilde{\nu}_5}}^2\tilde{N}_{5}^*\tilde{N}_{5}-m_{{\tilde{e}_5}}^2\tilde{E}_{5}^*\tilde{E}_{5}
-m_{{\Phi_{B}}}^2\Phi_{B}^*\Phi_{B}
\nonumber\\
&&\hspace{1.3cm}
-m_{{\varphi_{B}}}^2\varphi_{B}^*\varphi_{B}-m_{{\Phi_{L}}}^2\Phi_{L}^*\Phi_{L}
-m_{{\varphi_{L}}}^2\varphi_{L}^*\varphi_{L}-\Big(m_{B}\lambda_{B}\lambda_{B}
+m_{L}\lambda_{L}\lambda_{L}+h.c.\Big)
\nonumber\\
&&\hspace{1.3cm}
+\Big\{A_{{u_4}}Y_{{u_4}}\tilde{Q}_{4}H_{u}\tilde{U}_{4}^c+A_{{d_4}}Y_{{d_4}}\tilde{Q}_{4}H_{d}\tilde{D}_{4}^c
+A_{{u_5}}Y_{{u_5}}\tilde{Q}_{5}^cH_{d}\tilde{U}_{5}+A_{{d_5}}Y_{{d_5}}\tilde{Q}_{5}^cH_{u}\tilde{D}_{5}
\nonumber\\
&&\hspace{1.3cm}
+A_{{BQ}}\lambda_{Q}\tilde{Q}_{4}\tilde{Q}_{5}^c\Phi_{B}+A_{{BU}}\lambda_{U}\tilde{U}_{4}^c\tilde{U}_{5}\varphi_{B}
+A_{{BD}}\lambda_{D}\tilde{D}_{4}^c\tilde{D}_{5}\varphi_{B}+B_{B}\mu_{B}\Phi_{B}\varphi_{B}
+h.c.\Big\}
\nonumber\\
&&\hspace{1.3cm}
+\Big\{A_{{e_4}}Y_{{e_4}}\tilde{L}_{4}H_{d}\tilde{E}_{4}^c+A_{{\nu_4}}Y_{{\nu_4}}\tilde{L}_{4}H_{u}\tilde{N}_{4}^c
+A_{{e_5}}Y_{{e_5}}\tilde{L}_{5}^cH_{u}\tilde{E}_{5}+A_{{\nu_5}}Y_{{\nu_5}}\tilde{L}_{5}^cH_{d}\tilde{N}_{5}
\nonumber\\
&&\hspace{1.3cm}
+A_{N}Y_{\nu}\tilde{L}H_{u}\tilde{N}^c+A_{{N^c}}\lambda_{{N^c}}\tilde{N}^c\tilde{N}^c\varphi_{L}
+B_{L}\mu_{L}\Phi_{L}\varphi_{L}+h.c.\Big\}
\nonumber\\
&&\hspace{1.3cm}
+\Big\{A_1\lambda_1\tilde{Q}\tilde{Q}_{5}^cX+A_2\lambda_2\tilde{U}^c\tilde{U}_{5}X^\prime
+A_3\lambda_3\tilde{D}^c\tilde{D}_{5}X^\prime+B_{X}\mu_{X}XX^\prime+h.c.\Big\}\;.
\label{soft-breaking}
\end{eqnarray}

The $SU(2)_L$ doublets $H_{u},\;H_{d}$ should obtain nonzero VEVs $\upsilon_{u},\;\upsilon_{d}$,
\begin{eqnarray}
&&H_{u}=\left(\begin{array}{c}H_{u}^+\\{1\over\sqrt{2}}\Big(\upsilon_{u}+H_{u}^0+iP_{u}^0\Big)\end{array}\right)\;,~~~~
H_{d}=\left(\begin{array}{c}{1\over\sqrt{2}}\Big(\upsilon_{d}+H_{d}^0+iP_{d}^0\Big)\\H_{d}^-\end{array}\right)\;.
\end{eqnarray}
 The $SU(2)_L$ singlets $\Phi_{B},\;\varphi_{B}$ obtain nonzero VEVs
 $\upsilon_{{B}},\;\overline{\upsilon}_{{B}}$,
  \begin{eqnarray}
&&\Phi_{B}={1\over\sqrt{2}}\Big(\upsilon_{B}+\Phi_{B}^0+iP_{B}^0\Big)\;,~~~~~~~~~
\varphi_{B}={1\over\sqrt{2}}\Big(\overline{\upsilon}_{B}+\varphi_{B}^0+i\overline{P}_{B}^0\Big)\;.
\end{eqnarray}
In the same way, the $SU(2)_L$ singlets $\Phi_{L},\;
\varphi_{L}$  obtain nonzero VEVs $\upsilon_{L},\;\overline{\upsilon}_{L}$,
\begin{eqnarray}
&&\Phi_{L}={1\over\sqrt{2}}\Big(\upsilon_{L}+\Phi_{L}^0+iP_{L}^0\Big)\;,~~~~~~~~~~
\varphi_{L}={1\over\sqrt{2}}\Big(\overline{\upsilon}_{L}+\varphi_{L}^0+i\overline{P}_{L}^0\Big)\;.
\label{VEVs}
\end{eqnarray}
Therefore, the local gauge symmetry $SU(2)_{L}\otimes U(1)_{Y}\otimes U(1)_{B}\otimes U(1)_{L}$
breaks down to the electromagnetic symmetry $U(1)_{e}$.

$H^\pm=-\sin\beta H_{_d}^\pm+\cos\beta H_{_u}^\pm$
represent the charged Higgs, whose squared masses at tree level are
$m_{_{H^\pm}}^2=m_{_{A^0}}^2+m_{_{\rm W}}^2$.
The charged Goldstone bosons and neutral Goldstone bosons are denoted respectively
\begin{eqnarray}
&&G^\pm=\cos\beta H_{_d}^\pm+\sin\beta H_{_u}^\pm,~~~~~~ G^0=\cos\beta P_{_d}^0+\sin\beta P_{_u}^0\;,\nonumber\\
&&G_{_B}^0=\cos\beta_{_B} P_{_B}^0+\sin\beta_{_B}\overline{P}_{_B}^0, ~~~~G_{_L}^0=\cos\beta_{_L} P_{_L}^0+\sin\beta_{_L}\overline{P}_{_L}^0,
\label{Goldstone2}
\end{eqnarray}
with $\tan\beta=\upsilon_{_u}/\upsilon_{_d},\;
\tan\beta_{_B}=\overline{\upsilon}_{_B}/\upsilon_{_B},\;\tan\beta_{_L}=\overline{\upsilon}_{_L}/\upsilon_{_L}$.
\begin{eqnarray}
&&A^0=-\sin\beta P_{_d}^0+\cos\beta P_{_u}^0,~~A_{_B}^0=-\sin\beta_{_B} P_{_B}^0+\cos\beta_{_B}\overline{P}_{_B}^0\;,\nonumber\\
&&A_{_L}^0=-\sin\beta_{_L} P_{_L}^0+\cos\beta_{_L}\overline{P}_{_L}^0,
\label{neutral-pseudoscalar}
\end{eqnarray}
are the physical neutral pseudoscalar fields, and their masses at tree level read as\cite{weBLMSSM}
\begin{eqnarray}
&&m_{_{A^0}}^2={B\mu_{_H}\over\cos\beta\sin\beta},~~~m_{_{A_{_B}^0}}^2={B_{_B}\mu_{_B}\over\cos\beta_{_B}\sin\beta_{_B}},
~~~m_{_{A_{_L}^0}}^2={B_{_L}\mu_{_L}\over\cos\beta_{_L}\sin\beta_{_L}}.
\label{pseudoscalar-mass}
\end{eqnarray}

The lightest neutral CP-even Higgs $h^0$ is obtained from diagonalizing
the mass squared matrix of neutral CP-even Higgs in the sector ($H_{d}^0$,$H_{u}^0$)
\begin{eqnarray}
&&\left(\begin{array}{l}H^0\\h^0\end{array}\right)=\left(\begin{array}{cc}\cos\alpha&\sin\alpha\\
-\sin\alpha&\cos\alpha\end{array}\right)\left(\begin{array}{l}H_{_d}^0\\H_{_u}^0\end{array}\right),
\nonumber\\&&
\tan2\alpha=\frac{m_Z^2+m_{A^0}^2}{m_Z^2-m_{A^0}^2}\tan2\beta\;.
\label{charged-Higgs}
\end{eqnarray}

$\Phi_{_B}^0$ and $\varphi_{_B}^0$ mix together and the mass squared matrix is
\begin{eqnarray}
&&{\cal M}_{_{EB}}^2=\left(\begin{array}{ll}m_{_{Z_B}}^2\cos^2\beta_{_B}+m_{_{A_{_B}^0}}^2\sin^2\beta_{_B},\;&
(m_{_{Z_B}}^2+m_{_{A_{_B}^0}}^2)\cos\beta_{_B}\sin\beta_{_B}\\
(m_{_{Z_B}}^2+m_{_{A_{_B}^0}}^2)\cos\beta_{_B}\sin\beta_{_B},\;&
m_{_{Z_B}}^2\sin^2\beta_{_B}+m_{_{A_{_B}^0}}^2\cos^2\beta_{_B}
\end{array}\right)\;,
\label{CPevenB-mass}
\end{eqnarray}
with $v_{B_t}=\sqrt{\upsilon_{_B}^2+\overline{\upsilon}_{_B}^2}$. $m_{_{Z_B}}=g_{_B}v_{B_t}$ denotes the mass of
neutral $U(1)_{_B}$ gauge boson $Z_{_B}$.
 Two mass eigenstates can be gotten
\begin{eqnarray}
&&\left(\begin{array}{l}H_{_B}^0\\ h_{_B}^0\end{array}\right)=
\left(\begin{array}{cc}\cos\alpha_{_B}&\sin\alpha_{_B}\\-\sin\alpha_{_B}&\cos\alpha_{_B}\end{array}\right)
\left(\begin{array}{l}\Phi_{_B}^0\\ \varphi_{_B}^0\end{array}\right)\;,
\label{Mass-eigenstates-B}
\end{eqnarray}
by the  mixing angle $\alpha_{_B}$ that is defined as
\begin{eqnarray}
&&\tan2\alpha_{_B}={m_{_{Z_B}}^2+m_{_{A_{_B}^0}}^2\over m_{_{Z_B}}^2-m_{_{A_{_B}^0}}^2}
\tan2\beta_{_B}.
\label{Mixing-B}
\end{eqnarray}
In the same way, we obtain the mass squared matrix for $(\Phi_{_L}^0,\;\varphi_{_L}^0)$
\begin{eqnarray}
&&{\cal M}_{_{EL}}^2=\left(\begin{array}{ll}m_{_{Z_L}}^2\cos^2\beta_{_L}+m_{_{A_{_L}^0}}^2\sin^2\beta_{_L},\;&
(m_{_{Z_L}}^2+m_{_{A_{_L}^0}}^2)\cos\beta_{_L}\sin\beta_{_L}\\
(m_{_{Z_L}}^2+m_{_{A_{_L}^0}}^2)\cos\beta_{_L}\sin\beta_{_L},\;&
m_{_{Z_L}}^2\sin^2\beta_{_L}+m_{_{A_{_L}^0}}^2\cos^2\beta_{_L}
\end{array}\right),
\label{CPevenL-mass}
\end{eqnarray}
with $v_{L_t}=\sqrt{\upsilon_{_L}^2+\overline{\upsilon}_{_L}^2}$. $ m_{_{Z_L}}=2g_{_L}v_{L_t}$
 represents the mass of
neutral $U(1)_{_L}$ gauge boson $Z_{_L}$. In BLMSSM, the authors\cite{weBLMSSM,BLMass} analyze the mass matrices of exotic quarks,
exotic squarks and some exotic sleptons.

With the introduced superfields $\hat{N}^c$, three neutrinos obtain tiny masses through the see-saw
mechanism.
After symmetry breaking, we obtain the mass
matrix for neutrinos in the basis $(\nu,N^c)$\cite{BiaoChen}
    \begin{eqnarray}
\left(\begin{array}{cc}
  0&\frac{v_u}{\sqrt{2}}(Y_{\nu})^{IJ} \\
   \frac{v_u}{\sqrt{2}}(Y^{T}_{\nu})^{IJ}  & \frac{\bar{v}_L}{\sqrt{2}}(\lambda_{N^c})^{IJ}
    \end{array}\right)\label{numassM}.
      \end{eqnarray}

 Eq.(\ref{numassM}) can be diagonalized by the unitary matrix $U_\nu$.
Then, one gets three light and three heavy neutrino mass eigenstates.

 The new gaugino $\lambda_L$ and the superpartners of the $SU(2)_L$ singlets $\Phi_L,\varphi_L$ mix,
which produce three lepton neutralinos in the base $(i\lambda_L,\psi_{\Phi_L},\psi_{\varphi_L})$\cite{BLLEDM}.
\begin{equation}
\mathcal{L}_{\chi_L^0}=\frac{1}{2}(i\lambda_L,\psi_{\Phi_L},\psi_{\varphi_L})\left(     \begin{array}{ccc}
  2M_L &2v_Lg_L &-2\bar{v}_Lg_L\\
   2v_Lg_L & 0 &-\mu_L\\-2\bar{v}_Lg_L&-\mu_L &0
    \end{array}\right)  \left( \begin{array}{c}
 i\lambda_L \\ \psi_{\Phi_L}\\\psi_{\varphi_L}
    \end{array}\right)+h.c.\label{LN}
   \end{equation}
Using $Z_{N_L}$, one can diagonalize the mass matrix in Eq.(\ref{LN}) to obtain three lepton neutralino masses.

In BLMSSM, the mass squared matrix of slepton  is
different from that in MSSM, because of the contributions from Eqs.(\ref{superpotential-BL},\ref{soft-breaking}).
The corrected  mass squared matrix of slepton reads as
\begin{eqnarray}
&&\left(\begin{array}{cc}
  (\mathcal{M}^2_L)_{LL}&(\mathcal{M}^2_L)_{LR} \\
   (\mathcal{M}^2_L)_{LR}^{\dag} & (\mathcal{M}^2_L)_{RR}
    \end{array}\right).
\end{eqnarray}

$(\mathcal{M}^2_L)_{LL},~(\mathcal{M}^2_L)_{LR}$ and $(\mathcal{M}^2_L)_{RR}$ are shown here
\begin{eqnarray}
 &&(\mathcal{M}^2_L)_{LL}=\frac{(g_1^2-g_2^2)(v_d^2-v_u^2)}{8}\delta_{IJ} +g_L^2(\bar{v}_L^2-v_L^2)\delta_{IJ}
 +m_{l^I}^2\delta_{IJ}+(m^2_{\tilde{L}})_{IJ},\nonumber\\&&
 (\mathcal{M}^2_L)_{LR}=\frac{\mu^*v_u}{\sqrt{2}}(Y_l)_{IJ}-\frac{v_u}{\sqrt{2}}(A'_l)_{IJ}+\frac{v_d}{\sqrt{2}}(A_l)_{IJ},
 \nonumber\\&& (\mathcal{M}^2_L)_{RR}=\frac{g_1^2(v_u^2-v_d^2)}{4}\delta_{IJ}-g_L^2(\bar{v}_L^2-v_L^2)\delta_{IJ}
 +m_{l^I}^2\delta_{IJ}+(m^2_{\tilde{R}})_{IJ}.
\end{eqnarray}
The  unitary matrix $Z_{\tilde{L}}$ is used to rotate slepton mass squared matrix to mass eigenstates.

There are six sneutrinos, whose mass squared matrix is deduced
 from the superpotential and the soft breaking terms in Eqs.(\ref{superpotential-BL},\ref{soft-breaking}).
 In the base $\tilde{n}^{T}=(\tilde{\nu},\tilde{N}^{c})$,
  the concrete forms for the sneutrino mass squared matrix ${\cal M}_{\tilde{n}}$ are shown here
    \begin{eqnarray}
  && {\cal M}^2_{\tilde{n}}(\tilde{\nu}_{I}^*\tilde{\nu}_{J})=\frac{g_1^2+g_2^2}{8}(v_d^2-v_u^2)\delta_{IJ}+g_L^2(\overline{v}^2_L-v^2_L)\delta_{IJ}
   +\frac{v_u^2}{2}(Y^\dag_{\nu}Y_\nu)_{IJ}+(m^2_{\tilde{L}})_{IJ},\nonumber\\&&
   {\cal M}^2_{\tilde{n}}(\tilde{N}_I^{c*}\tilde{N}_J^c)=-g_L^2(\overline{v}^2_L-v^2_L)\delta_{IJ}
   +\frac{v_u^2}{2}(Y^\dag_{\nu}Y_\nu)_{IJ}+2\overline{v}^2_L(\lambda_{N^c}^\dag\lambda_{N^c})_{IJ}\nonumber\\&&
   \hspace{1.8cm}+(m^2_{\tilde{N}^c})_{IJ}+\mu_L\frac{v_L}{\sqrt{2}}(\lambda_{N^c})_{IJ}
   -\frac{\overline{v}_L}{\sqrt{2}}(A_{N^c})_{IJ}(\lambda_{N^c})_{IJ},\nonumber\\&&
   {\cal M}^2_{\tilde{n}}(\tilde{\nu}_I\tilde{N}_J^c)=\mu^*\frac{v_d}{\sqrt{2}}(Y_{\nu})_{IJ}-v_u\overline{v}_L(Y_{\nu}^\dag\lambda_{N^c})_{IJ}
   +\frac{v_u}{\sqrt{2}}(A_{N})_{IJ}(Y_\nu)_{IJ}.
   \end{eqnarray}

%\subsection{the couplings}
The superfields $\tilde{N}^c$ in BLMSSM lead to the corrections
for the couplings existed in MSSM and some corrected couplings are deduced.
We give out the couplings for W-lepton-neutrino and Z-neutrino-neutrino
\begin{eqnarray}
&&\mathcal{L}_{WL\nu}=-\frac{e}{\sqrt{2}s_W}W^+_{\mu}\sum_{I,J=1}^3\sum_{i=1}^2(U_{\nu^{IJ}}^\dag)^{i1}
\bar{\nu}^{I}_i\gamma^{\mu}P_Le^J,
\nonumber\\&&
\mathcal{L}_{Z\nu\nu}=-\frac{e}{2s_Wc_W}Z_{\mu}\sum_{I,J,K=1}^3\sum_{i,j=1}^2
(U_{\nu^{IK}}^\dag)^{i1}U_{\nu^{JK}}^{1j}\bar{\nu}^{I}_i\gamma^{\mu}P_L\nu^{J}_j.
\end{eqnarray}
The charged Higgs-lepton-neutrino couplings are
\begin{eqnarray}
&&\mathcal{L}_{H^{\pm}L\nu}=\sum_{I,J=1}^3\sum_{i=1}^2G^{\pm}
\bar{e}^J\Big(Y_l^{IJ}\cos\beta U_{\nu^{IJ}}^{1i}P_L-Y_{\nu}^{IJ*}\sin\beta U_{\nu^{IJ}}^{2i}P_R
\Big)\nu^I_i\nonumber\\&&-\sum_{I,J=1}^3\sum_{i=1}^2
H^{\pm}\bar{e}^J\Big(Y_l^{IJ}\sin\beta U_{\nu^{IJ}}^{1i}P_L+Y_{\nu}^{IJ*}\cos\beta U_{\nu^{IJ}}^{2i}P_R
\Big)\nu^I_i+h.c.\end{eqnarray}
The corrected chargino-lepton-sneutrino couplings read as
\begin{eqnarray}
&&\mathcal{L}_{\chi^{\pm}L\tilde{\nu}}=-\sum_{I,J=1}^3\sum_{i,j=1}^2\bar{\chi}^-_j
\Big(Y_l^{IJ} Z_-^{2j*}(Z_{\nu^{IJ}}^\dag)^{i1}P_R\nonumber\\&&+
[\frac{e}{s_W}Z_+^{1j}(Z_{\nu^{IJ}}^\dag)^{i1}+Y_\nu^{IJ}Z_+^{2j}(Z_{\nu^{IJ}}^\dag)^{i2}]P_L
\Big)e^J\tilde{\nu}^{I*}_i.\end{eqnarray}

 We also obtain the adapted Z-sneutrino-sneutrino couplings as follows
\begin{eqnarray}
\mathcal{L}_{Z\tilde{\nu}\tilde{\nu}}=-\frac{e}{2s_Wc_W}Z_{\mu}\sum_{I,J,K=1}^3
\sum_{i,j=1}^2(Z_{\nu^{IK}}^\dag)^{i1}Z_{\nu^{JK}}^{1j}\tilde{\nu}^{I*}_i\texttt{i}(\overrightarrow{\partial}^{\mu}
-\overleftarrow{\partial}^{\mu})\tilde{\nu}^{J}_j.
\end{eqnarray}

In BLMSSM, there are new couplings that are deduced from
the interactions of gauge and matter multiplets
$ig\sqrt{2}T^a_{ij}(\lambda^a\psi_jA_i^*-\bar{\lambda}^a\bar{\psi}_iA_j)$.
After calculation, the lepton-slepton-lepton neutralino couplings are obtained
\begin{eqnarray}
&&\mathcal{L}_{l\chi_L^0\tilde{L}}=\sqrt{2}g_L\bar{\chi}_{L_j}^0\Big(Z_{N_L}^{1j}Z_{L}^{Ii}P_L
-Z_{N_L}^{1j*}Z_{L}^{(I+3)i}P_R\Big)l^I\tilde{L}_i^++h.c.
\end{eqnarray}

\section{Charged Lepton flavor violation in the BLMSSM}
In this section, the CLFV processes $l_j\rightarrow l_i+\gamma$ and $l_j \rightarrow 3l_i$
are studied in the BLMSSM. For convenience, the triangle, penguin and box diagrams are analyzed in the generic form, which can simplify
the work.

\subsection{Rare decays $l_j\rightarrow l_i+\gamma$}

\begin{figure}[h]
\setlength{\unitlength}{1mm}
\centering
\includegraphics[width=5.6in]{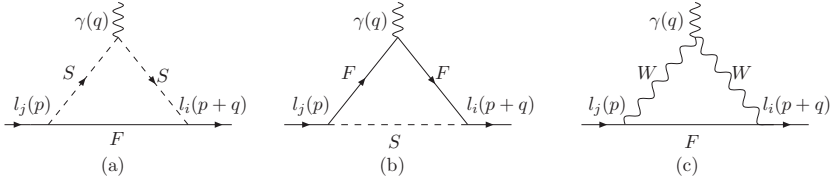}
\vspace{-16.0cm}
\caption[]{The one loop diagrams for $l_j\rightarrow l_i+\gamma$, with $F$ representing Dirac(Majorana) particles.}\label{FMFLV1}
\end{figure}

When the external leptons are all on shell, we can generally write the amplitudes for $l_j\rightarrow l_i+\gamma$
 as
\begin{eqnarray}
&&\mathcal{M} = e{\epsilon ^\mu }{\bar u_i}(p + q)\Big[q^2{\gamma _\mu }(C_1^LP_L + C_1^RP_R)
\nonumber\\
&&\qquad + \: {m_{{l_j}}}i{\sigma _{\mu \nu }}{q^\nu }(C_2^LP_L + C_2^RP_R)\Big]{u_j}(p)\:,
\label{amplitude-gamma}
\end{eqnarray}
where $p$ is the injecting lepton momentum, $q$ is the photon momentum, and $m_{{l_j}}$ is the mass of the $j$-th generation charged lepton.
  ${u_i}(p)$ and ${v_i}(p)$  are the wave functions for the external leptons.
The relevant Feynman diagrams are shown in Fig.\ref{FMFLV1}. The final Wilson coefficients
$C_1^L,C_1^R,C_2^L,C_2^R$ are obtained from the sum of these diagrams' amplitudes.

The contributions from the virtual neutral fermion diagrams Fig.\ref{FMFLV1}(a) are denoted by $C_\alpha^{L,R}(n),\alpha=1,2$.
We give out the deduced results in the following form,
\begin{eqnarray}
&&C_1^{L}(n) = \sum_{F=\chi^0,\chi_L^0,\nu}\sum_{S=\tilde{L},\tilde{L},H^{\pm}}\frac{1}{6{m_W^2}}H_R^{SF\bar{l}_i}H_L^{S^*l_j\bar{F}}{I_1}
({x_F},{x_S})\;,\nonumber\\
&&C_2^{L}(n) =\sum_{F=\chi^0,\chi_L^0,\nu}\sum_{S=\tilde{L},\tilde{L},H^{\pm}} \frac{{{m_F}}}{{{m_{{l_j}}}}{m_W^2}}H_L^{SF\bar{l}_i}H_L^{S^*l_j\bar{F}}
\Big[ {I_2}({x_F},{x_S}) - {I_3}({x_F},{x_S}) \Big]
\;, \nonumber\\
&&C_\alpha^{R}(n) = \left. {C_\alpha^{L}}(n) \right|{ _{L \leftrightarrow R}},~~~\alpha=1,2,\label{oneloopn}
\end{eqnarray}
with $x= {m^2}/{m_W^2}$ and $m$ representing the mass for the corresponding particle. $H_{L,R}^{SF\bar{l}_i}$ and $H_{L,R}^{S^*l_j\bar{F}}$
are the corresponding couplings of the left(right)-hand parts in the Lagrangian.
 The one-loop functions $I_{i}(x_1,x_2),i=1\dots3$ are collected here
\begin{eqnarray}
&&{I_1}(\textit{x}_1 , x_2 ) = \frac{1}{{96{\pi ^2}}} \Big[ \frac{{11 + 6\ln {x_2}}}{{({x_2} - {x_1})}}- \frac{{15{x_2} + 18{x_2}\ln {x_2}}}{{{{({x_2} - {x_1})}^2}}} + \frac{{6x_2^2 + 18x_2^2\ln {x_2}}}{{{{({x_2} - {x_1})}^3}}},  \nonumber\\
&&\qquad\qquad\quad\; + \: \frac{{6x_1^3\ln {x_1}}-{6x_2^3\ln {x_2}}}{{{{({x_2} - {x_1})}^4}}}  \Big]\:.\\
&&{I_2}(\textit{x}_1 , x_2 ) = \frac{1}{{32{\pi ^2}}}\Big[  \frac{{3 + 2\ln {x_2}}}{{({x_2} - {x_1})}} - \frac{{2{x_2} + 4{x_2}\ln {x_2}}}{{{{({x_2} - {x_1})}^2}}} -\frac{{2x_1^2\ln {x_1}}}{{{{({x_2} - {x_1})}^3}}}+ \frac{{2x_2^2\ln {x_2}}}{{{{({x_2} - {x_1})}^3}}}\Big]\:, \\&&{I_3}(\textit{x}_1 , x_2 ) = \frac{1}{{16{\pi ^2}}}\Big[ \frac{{1 + \ln {x_2}}}{{({x_2} - {x_1})}} + \frac{{{x_1}\ln {x_1}}-{{x_2}\ln {x_2}}}{{{{({x_2} - {x_1})}^2}}} \Big]\:.
\end{eqnarray}

  $C_\alpha^{L,R}(c),\alpha=1,2$ stand for the  coefficients from the virtual charged fermion diagrams Fig.\ref{FMFLV1}(b),
  and they are shown here
\begin{eqnarray}
&&C_1^{L}(c) = \sum_{F=\chi^{\pm}}\sum_{S=\tilde{\nu}} \frac{1}{6{m_W^2}}
H_R^{SF\bar{l}_i}H_L^{S^*l_j \bar{F}}
\Big[ {I_3}({x_F},{x_S})- 2 {I_4}({x_F},{x_S})- {I_1}({x_F},{x_S}) \Big] \;, \nonumber\\
&&C_2^{L}(c) = \sum_{F=\chi^{\pm}}\sum_{S=\tilde{\nu}} \frac{{{m_F}}}{{{m_{{l_j}}}}{m_W^2}}
H_L^{SF\bar{l}_i}H_L^{S^*l_j \bar{F}}\Big[ {I_3}({x_F},{x_S}) - {I_4}({x_F},{x_S}) - {I_1}({x_F},{x_S}) \Big] \;, \nonumber\\
&&C_\alpha^{R}(c) = \left. {C_\alpha^{L}}(c) \right|{ _{L \leftrightarrow R}}, ~~~\alpha=1,2.\label{oneloopc}
\end{eqnarray}
with
\begin{eqnarray}
{I_4}(\textit{x}_1 , x_2 ) = \frac{1}{{16{\pi ^2}}}\Big[ - \frac{{1 + \ln {x_1}}}{{({x_2} - {x_1})}} - \frac{{{x_1}\ln {x_1}}-{{x_2}\ln {x_2}}}{{{{({x_2} - {x_1})}^2}}} \Big]\:.
\end{eqnarray}

Because three light neutrinos and three heavy neutrinos mix together, the virtual $W$ diagrams Fig.\ref{FMFLV1}(c) have
corrections to the CLFV process $l_j\rightarrow l_i\gamma$. The corresponding coefficients are denoted by
$C_\alpha^{L,R}(W)(\alpha=1,2)$
\begin{eqnarray}
&&C^L_1(W)=\sum_{F=\nu}\frac{-1}{2m_W^2}H^{WF\bar{l}_i}_LH^{W^*l_j\bar{F}}_L\Big[I_2(x_{F},x_W)+I_1(x_{F},x_W)\Big],
\nonumber\\&&C^L_2(W)=\sum_{F=\nu}\frac{1}{m_W^2}H^{WF\bar{l}_i}_LH^{W^*l_j\bar{F}}_L(1+\frac{m_{l_i}}{m_{l_j}})
\Big[2I_2(x_{F},x_W)-\frac{1}{3}I_1(x_{F},x_W)\Big],
\nonumber\\&&C_\alpha^{R}(W)=0,~~~\alpha=1,2.\label{oneloopW}
\end{eqnarray}
The total coefficients are the sum of Eqs.(\ref{oneloopn})(\ref{oneloopc})(\ref{oneloopW})
\begin{eqnarray}
C_\alpha^{L,R}=C_\alpha^{L,R}(n)+C_\alpha^{L,R}(c)+C_\alpha^{L,R}(W),~~~i=1,2.
\end{eqnarray}
With the Eq.(\ref{amplitude-gamma}), the decay width for $l_j\rightarrow l_i+\gamma$ can be expressed as\cite{ljlig}
\begin{eqnarray}
\Gamma (l_j  \to l_i+\gamma ) = \frac{{{e^2}}}{{16\pi }}m_{{l_j}}^5 \Big({\left| {C_2^L} \right|^2} + {\left| {C_2^R} \right|^2}\Big)\:.
\label{gamma-1}
\end{eqnarray}

\subsection{Rare decays $l_j \rightarrow 3l_i$}
The CLFV processes $l_j  \rightarrow 3l_i$ are very interesting. Both penguin-type diagrams and box-type diagrams have contributions to the effective Lagrangian. With Eq.(\ref{amplitude-gamma}), one can obtain the $\gamma$-penguin contributions in the following form,
\begin{eqnarray}
&&T_{\gamma - {\rm{p}}} = {\bar u_i}({p_1})\Big[{q^2}{\gamma _\mu }(C_1^L{P_L}
+ C_1^RP_R) + {m_{l_j}}i{\sigma _{\mu \nu }}{q^\nu }(C_2^LP_L
+ C_2^RP_R)\Big] {u_j}(p)  \nonumber\\
&&\qquad\quad\;\; \times \: \frac{{{e^2}}}{{{q^2}}}{\bar u_i}({p_2}){\gamma ^\mu }
{v_i}({p_3})- ({p_1} \leftrightarrow {p_2})\:.
\end{eqnarray}

\begin{figure}[h]
\setlength{\unitlength}{1mm}
\centering
\includegraphics[width=2.0in]{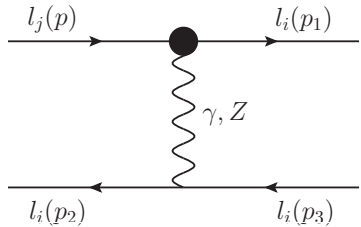}
\vspace{-0.0cm}
\caption[]{The penguin-type diagrams for CLFV process $l_j\rightarrow 3l_i$.}\label{FMFLV2}
\end{figure}

The contributions from $Z$-penguin diagrams are depicted by the Fig.\ref{FMFLV2}, and
deduced in the same way as $\gamma$-penguin diagrams,
\begin{eqnarray}
&&T_{Z- {\rm{p}}} = \frac{{{e^2}}}{{m_Z^2}}{\bar u_i}({p_1}){\gamma _\mu }
({N_L}{P_L} + {N_R}{P_R}){u_j}(p)  {\bar u_i}({p_2}){\gamma ^\mu }\Big(H_L^{Zl_i{\bar{l} _i}}{P_L} \qquad \nonumber\\
&&\qquad\quad\;\; + \:H_R^{Zl_i{\bar{l} _i}}{P_R}\Big){v_i}({p_3}) - ({p_1} \leftrightarrow {p_2})\:,
\nonumber\\&&{N_{L,R}} = N_{L,R}(S) +N_{L,R}(W)\:.
\end{eqnarray}

The concrete forms of the effective couplings $N_{L}(S),~N_{R}(S)$ read as
\begin{eqnarray}
&&N_L(S) = \frac{1}{2{e^2}}\sum\limits_{F=\chi^0,\chi^{\pm},\nu}
 \sum\limits_{S=\tilde{L},\tilde{\nu},H^{\pm}}\Big[\frac{2{m_{F_1}}{m_{F_2}}}
{{m_W^2}}H_R^{SF_2{{\bar l }_{i}}}H_L^{ZF_1{{\bar F }_2 }}
H_L^{S^*{l _{j}}{{\bar F }_1 }}{G_1}({x_S},{x_{F_2}},{x_{F_1}}) \nonumber\\&&\hspace{1.2cm}+H_R^{S_2 F{{\bar l }_{i}}}
H_R^{ZS_1 S_2^ {*} }H_L^{S_1 ^{*} {l _{j}}\bar F}{G_2}
({x_F},{x_{S_1 }},{x_{S_2}})-  H_R^{SF_2{{\bar l }_{i}}}
H_R^{ZF_1{{\bar F }_2 }}H_L^{S^*{l _{j}}{{\bar F }_1 }}{G_2}({x_S},{x_{F_2}},{x_{F_1}})
\Big]\nonumber\\&&\hspace{1.2cm}+\sum\limits_{F=\chi^0_L}
 \sum\limits_{S=\tilde{L}}\Big[H_R^{S_2 F{{\bar l }_{i}}}
H_R^{ZS_1 S_2^ {*} }H_L^{S_1 ^{*} {l _{j}}\bar F}{G_2}
({x_F},{x_{S_1 }},{x_{S_2}})\Big],\nonumber\\
&&N_R(S) = \left. {N_L}(S) \right|{ _{L \leftrightarrow R}} \:.
\end{eqnarray}
The functions $G_1(x_1,x_2,x_3)$ and $G_2(x_1,x_2,x_3)$ are
\begin{eqnarray}
&&{G_1}(\textit{x}_1 , x_2 , x_3) =  \frac{1}{{16{\pi ^2}}}\Big[ \frac{{{x_1}\ln {x_1}}}{{({x_1} - {x_2})({x_1} - {x_3})}} + \frac{{{x_2}\ln {x_2}}}{{({x_2} - {x_1})({x_2} - {x_3})}} \nonumber\\
&&\qquad\qquad\qquad\quad +  \: \frac{{{x_3}\ln {x_3}}}{{({x_3} - {x_1})({x_3} - {x_2})}}\Big], \\
&&{G_2}(\textit{x}_1 , x_2 , x_3) =  \frac{1}{{16{\pi ^2}}}\Big[ -(\Delta  + 1 + \ln {x_\mu })  + \frac{{x_1^2\ln {x_1}}}{{({x_1} - {x_2})({x_1} - {x_3})}}  \nonumber\\
&&\qquad\qquad\qquad\quad  + \: \frac{{x_2^2\ln {x_2}}}{{({x_2} - {x_1})({x_2} - {x_3})}}+ \frac{{x_3^2\ln {x_3}}}{{({x_3} - {x_1})({x_3} - {x_2})}} \Big].
\end{eqnarray}
${G_2}(\textit{x}_1 , x_2 , x_3)$ has infinite term, and to obtain finite results we use $\overline{MS}$ subtraction and $\overline{DR}$ scheme.

We deduce the effective couplings $N_{L,R}(W)$ in detail and keep the small $m_i$ terms.
\begin{eqnarray}
&&N_L(W)=\frac{c_W}{es_W}\sum_{F=\nu}H^{WF\bar{l}_i}_LH^{W^*l_j\bar{F}}_L
\Big[G_3(x_F,x_W)+2(x_i+x_j)[I_1(x_F,x_W)-I_2(x_F,x_W)]\Big]\nonumber\\&&\hspace{1.8cm}
+\frac{1}{e^2}\sum_{F_1,F_2=\nu}H^{WF_2\bar{l}_i}_LH^{W^*l_j\bar{F}_1}_LH^{Z^*F_1\bar{F}_2}_L
\Big(-\frac{3}{32\pi^2}-G_2(x_W,x_{F_1},x_{F_2})\nonumber\\&&\hspace{1.8cm}+x_j[\frac{1}{3}G_4(x_W,x_{F_1},x_{F_2})
+G_5(x_W,x_{F_1},x_{F_2})]
\Big),\nonumber\\&&
N_R(W)=\frac{c_W}{es_W}\sum_{F=\nu}H^{WF\bar{l}_i}_LH^{W^*l_j\bar{F}}_L\Big[2\sqrt{x_ix_j}[I_1(x_F,x_W)-I_2(x_F,x_W)]\Big]
\nonumber\\&&\hspace{1.8cm}+\frac{1}{e^2}\sum_{F_1,F_2=\nu}H^{WF_2\bar{l}_i}_LH^{W^*l_j\bar{F}_1}_LH^{Z^*F_1\bar{F}_2}_L\sqrt{x_ix_j}
\Big(2G_1(x_W,x_{F_1},x_{F_2})\nonumber\\&&\hspace{1.8cm}
-\frac{1}{3}G_4(x_W,x_{F_1},x_{F_2})
-2G_5(x_W,x_{F_1},x_{F_2})  \Big).
\end{eqnarray}
 The concrete expressions for the functions $G_3(x_1,x_2),G_4(x_1,x_2,x_3)$ and $G_5(x_1,x_2,x_3)$ are collected here
\begin{eqnarray}
&&G_3(x_1,x_2)=\frac{-1}{16 \pi ^2}\Big((\Delta+\ln
   x_\mu+1)+\frac{x_2^2
   \ln x_2-x_1^2 \ln x_1}{(x_2-x_1)^2}+\frac{x_2+2 x_2 \ln
   x_2}{x_1-x_2}-\frac{1}{2}\Big),\nonumber\\&&
G_4(x_1,x_2,x_3)=\frac{1}{32\pi^2}\Big(\frac{2x_1^3[3x_1(x_1-x_2-x_3)+x_2^2+x_2x_3+x_3^2]\ln x_1}{(x_1-x_2)^3(x_1-x_3)^3}\nonumber\\&&\hspace{2.6cm}-\frac{2(3x_1^2-3x_1x_2+x_2^2)x_2\ln x_2}{(x_1-x_2)^3(x_2-x_3)}+\frac{2(3x_1^2-3x_1x_3+x_3^2)x_3\ln x_3}{(x_1-x_3)^3(x_2-x_3)}\nonumber\\&&\hspace{2.6cm}
-\frac{x_1[5x_1^2-3x_1(x_2+x_3)+x_2x_3]}{(x_1-x_2)^2(x_1-x_3)^2}
\Big),\nonumber\\
&&G_5(x_1,x_2,x_3)=\frac{1}{16\pi^2}\Big(\frac{x_1^2(2x_1-x_2-x_3)\ln x_1}{(x_1-x_2)^2(x_1-x_3)^2}+\frac{x_2(x_2-2x_1)\ln x_2}{(x_1-x_2)^2(x_2-x_3)}
\nonumber\\&&\hspace{2.6cm}-\frac{x_1}{(x_1-x_2)(x_1-x_3)}+\frac{x_3(2x_1-x_3)\ln x_3}{(x_1-x_3)^2(x_2-x_3)}\Big).
\end{eqnarray}

\begin{figure}[h]
\setlength{\unitlength}{1mm}
\centering
\includegraphics[width=6.0in]{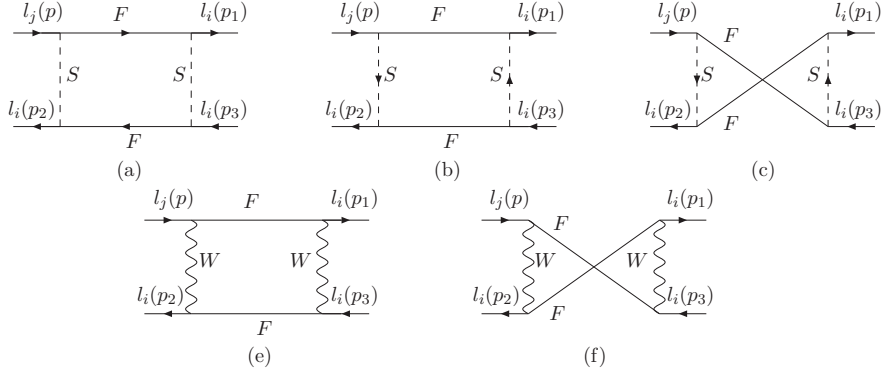}
\vspace{-15.0cm}
\caption[]{The box-type diagrams for CLFV processes $l_j\rightarrow 3l_i$ with $F$ representing Dirac(Majorana) particles.}\label{FMFLV3}
\end{figure}

The box-type diagrams drawn in Fig.\ref{FMFLV3}  can be written as
\begin{eqnarray}
&&T_{box} = \Big\{B_1^L{e^2}{\bar u_i}({p_1}){\gamma _\mu }{P_L}{u_j}(p){\bar u_i}(p_2)
{\gamma ^\mu }{P_L}{v_i}({p_3}) + (L \leftrightarrow R)\Big\}   \nonumber\\
&&\quad + \: \Big\{B_2^L{e^2}\Big[{\bar u_i}(p_1){\gamma _\mu }{P_L}{u_j}(p){\bar u_i}(p_2)
{\gamma ^\mu }{P_R}{v_i}({p_3}) - (p_1 \leftrightarrow {p_2})\Big] + {(L \leftrightarrow R) } \Big\} \nonumber\\
&&\quad + \: \Big\{ B_3^L{e^2}\Big[{\bar u_i}({p_1}){P_L}{u_j}(p){\bar u_i}({p_2})
{P_L}{v_i}({p_3}) - ({p_1} \leftrightarrow {p_2})\Big] + (L \leftrightarrow R) \Big\}   \nonumber\\
&&\quad + \: \Big\{ B_4^L{e^2}\Big[{\bar u_i}({p_1}){\sigma _{\mu \nu }}{P_L}{u_j}(p){\bar u_i}({p_2}){\sigma ^{\mu \nu }}{P_L}{v_i}({p_3})
- ({p_1} \leftrightarrow {p_2})\Big]  + \: (L \leftrightarrow R)\Big\} \:
\nonumber\\&&\quad + \: \Big\{ B_5^L{e^2}\Big[{\bar u_i}({p_1}){P_L}{u_j}(p){\bar u_i}({p_2}){P_R}{v_i}({p_3})
- ({p_1} \leftrightarrow {p_2})\Big]  + \: (L \leftrightarrow R)\Big\}.
\end{eqnarray}

From the box-type diagrams, we obtain the virtual chargino contributions to the effective couplings $B_\beta^{L,R}(c)$ with $\beta=1\dots5$
\begin{eqnarray}
&&B_1^{L}(c) = \sum_{F_1,F_2=\chi^{\pm}} \sum_{S_1,S_2=\tilde{\nu}}\frac{1}{2{e^2}{m_W^2}}
{G_6}(x_{F_1},x_{F_2},x_{S_1},x_{S_2})H_R^{S_2F_1{{\bar l }_{i}}}
H_L^{S_1l_j{{\bar F }_1}}H_R^{S_1F_2{{\bar l }_{i}}}
H_L^{S_2{l _{i}}{{\bar F }_2 }},\nonumber\\
&&B_2^{L}(c) = \sum_{F_1,F_2=\chi^{\pm}} \sum_{S_1,S_2=\tilde{\nu}} \Big[ \frac{1}{4{e^2}{m_W^2}} {G_6}(x_{F_1},x_{F_2},x_{S_1},x_{S_2})
H_R^{S_2F_1{{\bar l }_{i}}}
H_L^{S_1l_j{{\bar F }_1}}H_L^{S_1F_2{{\bar l }_{i}}}
H_R^{S_2{l _{i}}{{\bar F }_2 }}\nonumber\\
&&\qquad\;\;\quad - \frac{m_{F_1}
m_{F_2}}{2{e^2}{m_W^4}}
{G_7}(x_{F_1},x_{F_2},x_{S_1},x_{S_2})H_R^{S_2F_1{{\bar l }_{i}}}
H_R^{S_1l_j{{\bar F }_1}}H_L^{S_1F_2{{\bar l }_{i}}}
H_L^{S_2{l _{i}}{{\bar F }_2 }} \Big] ,
\nonumber\\
&&B_3^{L}(c) = \sum_{F_1,F_2=\chi^{\pm}} \sum_{S_1,S_2=\tilde{\nu}} \frac{m_{F_1}
m_{F_2}}{{e^2}{m_W^4}} {{G_7}(x_{F_1},x_{F_2},x_{S_1},x_{S_2})
H_L^{S_2F_1{{\bar l }_{i}}}
H_L^{S_1l_j{{\bar F }_1}}H_L^{S_1F_2{{\bar l }_{i}}}
H_L^{S_2{l _{i}}{{\bar F }_2 }}},
\nonumber\\
&&B_4^{L} (c)= B_5^{L} (c)= 0\:, ~~B_\beta^{R}(c) = \left. {B_\beta^{L}}(c) \right|{ _{L \leftrightarrow R}},
~~~\beta=1\dots5.\label{B4C}
\end{eqnarray}
with
\begin{eqnarray}
&&{G_6}(\textit{x}_1 , x_2 , x_3, x_4) = \frac{1}{{16{\pi ^2}}}\Big[\frac{{x_1^2\ln {x_1}}}{{({x_1} - {x_2})({x_1} - {x_3})({x_1} - {x_4})}}  +\; \frac{{x_2^2\ln {x_2}}}{{({x_2} - {x_1})({x_2} - {x_3})({x_2} - {x_4})}}\nonumber\\
&&\hspace{3.0cm}  + \frac{{x_3^2\ln {x_3}}}{{({x_3}  - {x_1})({x_3} - {x_2})({x_3} - {x_4})}} + \frac{{x_4^2\ln {x_4}}}{{({x_4} - {x_1})({x_4} - {x_2})({x_4} - {x_3})}}\Big]\:,
\nonumber\\&&
%%%%%%%%%%%%%%%%%%%%%%%%%%%%%%%%%%%%%%%%%%%%%%%%%%%%%%%%%%
{G_7}(\textit{x}_1 , x_2 , x_3, x_4) = \frac{1}{{16{\pi ^2}}}\Big[\frac{{{x_1}\ln {x_1}}}{{({x_1} - {x_2})({x_1} - {x_3})({x_1} - {x_4})}}  +\frac{{{x_2}\ln {x_2}}}{{({x_2} - {x_1})({x_2} - {x_3})({x_2} - {x_4})}} \nonumber\\
&&\hspace{3.2cm} + \frac{{{x_3}\ln {x_3}}}{{({x_3}  - {x_1})({x_3} - {x_2})({x_3} - {x_4})}}  + \: \frac{{{x_4}\ln {x_4}}}{{({x_4} - {x_1})({x_4} - {x_2})({x_4} - {x_3})}}\Big] .
\end{eqnarray}
For the box-type diagrams, the neutralino-slepton, neutrino-charged Higgs and lepton neutralino-slepton  contributions
to the effective couplings $B_\beta^{L,R}(n)$ are gotten,
\begin{eqnarray}
&&B_1^{L}(n) = \sum_{F_1,F_2=\chi^0,\chi^0_L,\nu}\sum_{S_1,S_2=\tilde{L},\tilde{L},H^{\pm}}\frac{{m_{F_1}}{m_{F_2}}}{{e^2}{m_W^4}}
{G_7}(x_{F_1},x_{F_2},x_{S_1},x_{S_2})H_L^{S_2 F_1{{\bar  l }_{i}}}
H_L^{S_1^*{l _{j}}\bar{F}_1}H_R^{S_2 F_2{{\bar l }_{i}}}
H_R^{S_1^* {l _{i}}\bar{F}_2}
\nonumber\\
&&\qquad\;\quad +\:\frac{1}{2{e^2}{m_W^2}}{G_6}(x_{F_1},x_{F_2},x_{S_1},x_{S_2})\Big[
H_R^{S_2 F_1{{\bar  l }_{i}}}
H_L^{S_1^*{l _{j}}\bar{F}_1}H_R^{S_1 F_2{{\bar l }_{i}}}
H_L^{S_2^* {l _{i}}\bar{F}_2}
\nonumber\\
&&\qquad\;\quad + \:H_L^{S_2 F_1{{\bar  l }_{i}}}
H_R^{S_1^*{l _{j}}\bar{F}_1}H_R^{S_2 F_2{{\bar l }_{i}}}
H_L^{S_1^* {l _{i}}\bar{F}_2}\Big]  ,
\nonumber\\
&&B_2^{L}(n) =\sum_{F_1,F_2=\chi^0,\chi^0_L,\nu}\sum_{S_1,S_2=\tilde{L},\tilde{L},H^{\pm}} - \frac{{m_{F_1}}{m_{F_2}}}
{2{e^2}{m_W^4}}{G_7}(x_{F_1},x_{F_2},x_{S_1},x_{S_2})H_R^{S_2 F_1{{\bar  l }_{i}}}
H_R^{S_1^*{l _{j}}\bar{F}_1}H_L^{S_1 F_2{{\bar l }_{i}}}H_L^{S_2^* {l _{i}}\bar{F}_2}
\nonumber\\
&&\qquad\;\quad +\:\frac{1}{4{e^2}{m_W^2}}{G_6}(x_{F_1},x_{F_2},x_{S_1},x_{S_2})\Big[
 H_R^{S_2 F_1{{\bar  l }_{i}}}
H_L^{S_1^*{l _{j}}\bar{F}_1}H_L^{S_1 F_2{{\bar l }_{i}}}H_R^{S_2^* {l _{i}}\bar{F}_2}
\nonumber\\
&&\qquad\;\quad + \: H_R^{S_2 F_1{{\bar  l }_{i}}}
H_L^{S_1^*{l _{j}}\bar{F}_1}H_R^{S_2 F_2{{\bar l }_{i}}}H_L^{S_1^* {l _{i}}\bar{F}_2}\Big]   ,
\nonumber\\
&&B_3^{L}(n)=\sum_{F_1,F_2=\chi^0,\chi^0_L,\nu}
\sum_{S_1,S_2=\tilde{L},\tilde{L},H^{\pm}}\frac{{m_{F_1}}{m_{F_2}}}{{e^2}{m_W^4}}{G_7}(x_{F_1},x_{F_2},x_{S_1},x_{S_2})\Big[
 H_L^{S_2 F_1{{\bar  l }_{i}}}
H_L^{S_1^*{l _{j}}\bar{F}_1}H_L^{S_1 F_2{{\bar l }_{i}}}H_L^{S_2^* {l _{i}}\bar{F}_2}
\nonumber\\
&&\qquad\;\quad - \:\frac{1}{2}\: H_L^{S_2 F_1{{\bar  l }_{i}}}
H_L^{S_1^*{l _{j}}\bar{F}_1}H_L^{S_2 F_2{{\bar l }_{i}}}H_L^{S_1^* {l _{i}}\bar{F}_2}\Big] ,
\nonumber\\
&&B_4^{L}(n)=\sum_{F_1,F_2=\chi^0,\chi^0_L,\nu}\sum_{S_1,S_2=\tilde{L},\tilde{L},H^{\pm}}
\frac{{m_{F_1}}{m_{F_2}}}{8{e^2}{m_W^4}}{G_7}(x_{F_1},x_{F_2},x_{S_1},x_{S_2})
H_L^{S_2 F_1{{\bar  l }_{i}}}
H_L^{S_1^*{l _{j}}\bar{F}_1}H_L^{S_2 F_2{{\bar l }_{i}}}H_L^{S_1^* {l _{i}}\bar{F}_2},
\nonumber\\&&
B_5^{L}(n)=0,
~~~~~B_\beta^{R}(n) = \left. {B_\beta^{L}}(n) \right|
{ _{L \leftrightarrow R}},~~~~~\beta=1\dots5.\label{B4N}
\end{eqnarray}
We also deduce the box-type contributions from virtual W-neutrino
\begin{eqnarray}
&&B_1^{L}(W) = \sum_{F_1,F_2=\nu}\frac{1}{{e^2}{m_W^2}}\Big[-\frac{\partial }{\partial x_W}
{G_2}(x_{W},x_{F_1},x_{F_2})H_L^{Wl_j\bar{F}_1}H_L^{W^*F_1\bar{l}_i}H_L^{W^*F_2\bar{l}_i}H_L^{Wl_i\bar{F}_2}\nonumber\\&&
-2\frac{m_{F_1}m_{F_2}}{{m_W^2}}\frac{\partial }{\partial x_W}
{G_1}(x_{W},x_{F_1},x_{F_2})H_L^{Wl_j\bar{F}_1}H_L^{W^*F_2\bar{l}_i}H_L^{Wl_i\bar{F}_2}H_L^{W^*F_1\bar{l}_i}\Big],
\nonumber\\&&
B_3^L(W)=\sum_{F_1,F_2=\nu}-\frac{7}{2}\frac{m_{F_1}m_{F_2}}{{e^2m_W^4}}\frac{\partial }{\partial x_W}
{G_1}(x_{W},x_{F_1},x_{F_2})H_L^{Wl_j\bar{F}_1}H_L^{W^*F_2\bar{l}_i}H_L^{Wl_i\bar{F}_2}H_L^{W^*F_1\bar{l}_i},\nonumber\\&&
B_2^L(W)=0,~~~B_4^L(W)=\frac{1}{14}B_3^L(W),~~B_5^L(W)=-\frac{1}{7}B_3^L(W),\nonumber\\&&
B_1^{R}(W)=B_2^{R}(W)=0,~~~
B_t^{R}(W) = {B_t^{L}}(W),~~~t=3,4,5.\label{B4W}
\end{eqnarray}

With Eqs.(\ref{B4C},\ref{B4N},\ref{B4W}), $B_\beta^{L,R}$ are expressed as
\begin{eqnarray}
B_\beta^{L,R} = B_\beta^{L,R}(n) + B_\beta^{L,R}(c)+ B_\beta^{L,R}(W),~~~(\beta = 1\dots5)\:.
\end{eqnarray}

The decay widths for $l_j  \rightarrow 3l_i$ can be computed
from the front amplitudes,
\begin{eqnarray}
&&\Gamma (l_j  \to 3l_i) = \frac{{{e^4}}}{{512{\pi ^3}}}m_{{l_j}}^5  \Big\{
({\left| {C_2^L} \right|^2} + {\left| {C_2^R} \right|^2})(\frac{{16}}{3}\ln
\frac{{{m_{{l_j}}}}}{{2{m_{{l_i}}}}} - \frac{{14}}{9}) \nonumber\\
&&\quad + \: ({\left| {C_1^L} \right|^2} + {\left| {C_1^R} \right|^2}) - 2(C_1^LC_2^{R * }
 + C_2^LC_1^{R * } + \textrm{H.c.}) + \frac{1}{6}({\left| {B_1^L} \right|^2}  + {\left| {B_1^R} \right|^2}) \nonumber\\
&&\quad  + \: \frac{1}{3}({\left| {B_2^L} \right|^2} + {\left| {B_2^R} \right|^2})
+ \frac{1}{{24}}({\left| {B_3^L} \right|^2} + {\left| {B_3^R} \right|^2}) + 6({\left| {B_4^L} \right|^2} + {\left| {B_4^R} \right|^2})
 \nonumber\\&&
\quad+\;\frac{1}{12}({\left| {B_5^L} \right|^2} + {\left| {B_5^R} \right|^2})
- \frac{1}{6}(B_2^LB_5^{L * } + B_2^RB_5^{R * }
+C_1^LB_5^{L * } + C_1^RB_5^{R * }+ \textrm{H.c.}) \nonumber\\&&\quad
+\;\frac{1}{3}(C_2^RB_5^{L * } + C_2^LB_5^{R * }+ \textrm{H.c.})- \frac{1}{6}(N_{LR}B_5^{L * } + N_{RL}B_5^{R * }
+ \textrm{H.c.})
\nonumber\\
&&\quad  - \: \frac{1}{2}(B_3^LB_4^{L * } + B_3^RB_4^{R * } + \textrm{H.c.}) +
\frac{1}{3}(C_1^LB_1^{L * } + C_1^RB_1^{R * } + C_1^LB_2^{L * }  \nonumber\\
&&\quad  + \: C_1^RB_2^{R * } + \textrm{H.c.}) - \frac{2}{3}(C_2^RB_1^{L * } + C_2^LB_1^{R * }  +
 C_2^LB_2^{R * } + C_2^RB_2^{L * } + \textrm{H.c.})  \nonumber\\
&&\quad + \: \frac{1}{3}\Big[2({\left| {{N_{LL}}} \right|^2} + {\left| {{N_{RR}}} \right|^2})
+ ({\left| {{N_{LR}}} \right|^2} + {\left| {{N_{RL}}} \right|^2}) + (B_1^LN_{LL}^ *  + B_1^RN_{RR}^ *   \nonumber\\
&&\quad  + \: B_2^LN_{LR}^ *  + B_2^RN_{RL}^ *  + \textrm{H.c.}) + 2(C_1^LN_{LL}^ *  + C_1^RN_{RR}^ *  + \textrm{H.c.})  \nonumber\\
&&\quad + \: (C_1^LN_{LR}^ *  + C_1^RN_{RL}^ *  + \textrm{H.c.}) - 4(C_2^RN_{LL}^ *  + C_2^LN_{RR}^ *  + \textrm{H.c.})  \nonumber\\
&&\quad  - \: 2(C_2^LN_{RL}^ *  + C_2^RN_{LR}^ *  + \textrm{H.c.})\Big]\Big\},
\label{gamma-2}
\end{eqnarray}
with
\begin{eqnarray}
&&{N_{LL}} = \frac{{{N_L}H_L^{Z{l_i}{\bar{l} _i}}}}{{m_Z^2}},\qquad
{N_{RR}} = {N_{LL}}\left| {_{L \leftrightarrow R}} \right., \quad \nonumber\\
&&{N_{LR}} = \frac{{{N_L}H_R^{Z{l _{ i}}{\bar{l} _{ i}}}}}{{m_Z^2}},\qquad
{N_{RL}} = {N_{LR}}\left| {_{L \leftrightarrow R}} \right.. \quad
\end{eqnarray}
With the fomula $\frac{\Gamma(l_j\rightarrow3l_i)}{\Gamma(l_j)}$, the branching ratios of $l_j \to 3l_i$ are obtained.
\section{numerical results}
In this section we discuss the numerical results, and consider the experiment constraints from
the lightest neutral CP-even Higgs mass
$m_{_{h^0}}\simeq125.7\;{\rm GeV}$ \cite{Higgs} and the neutrino experiment data. In this model, the neutron EDM, lepton EDM
and muon MDM are studied in our previous works, and their constraints are also taken into account.
In this work, we use the parameters\cite{BiaoChen} $L_4={3\over2},
\lambda_{N^c}=1$. The Yukawa couplings of neutrinos $(Y_\nu)^{IJ},(I,J=1,2,3)$ are at the order of $10^{-8}\sim10^{-6}$, whose effects to the
CLFV processes are tiny and can be ignored.

 To simplify the numerical discussion, we use the following relations
  \begin{eqnarray}
 &&(A_l)_{ii}=AL,~~~~(A_{N^c})_{ii}=(A_{N})_{ii}=AN,~~~(A'_l)_{ii}=A'_L,\nonumber\\&&
(m^2_{\tilde{N}^c})_{ii}=M_{sn}^2,
~~~(m_{\tilde{L}}^2)_{ii}=(m_{\tilde{R}}^2)_{ii}=s_m^2, ~~~\texttt{for}~~ i=1,2,3.
\end{eqnarray}
If we do not specially declare, the non-diagonal elements of the used parameters should be zero.

\subsection{$l_j\rightarrow l_i+\gamma$}

Charged lepton flavor violation is related to the new physics, and the branching ratio of the process
 $\mu\rightarrow e+\gamma$ is strict. Its experiment upper bound is
 $5.7\times10^{-13}$ at $90\%$ confidence level. At this subsection, the supposed parameters are $AN=-500{\rm GeV},m_L=3{\rm TeV}$.

1. $\mu\rightarrow e+\gamma$

\begin{figure}[h]
\setlength{\unitlength}{1mm}
\centering
\includegraphics[width=3.3in]{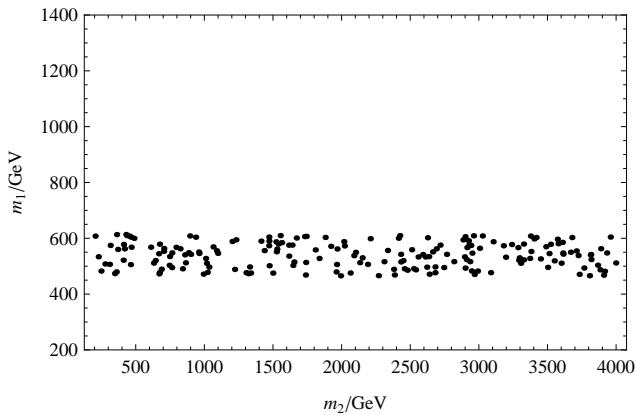}
\caption[]{For $\mu\rightarrow e+\gamma$, the allowed parameters in the plane of $m_1$ versus $m_2$ with
 $\mu_H=480{\rm GeV}, \tan\beta=11, s_m=3400{\rm GeV}, AL=-2{\rm TeV}, A^{'}_L=300{\rm GeV}$.}\label{muem1m2}
\end{figure}

With the parameters $M_{sn}=1{\rm TeV},\mu_L=1{\rm TeV}, g_L=1/6, v_{L_t}=3{\rm TeV},\tan\beta_L=2$, we numerically study the CLFV process
$\mu\rightarrow e+\gamma$. The mass matrix of neutralino includes $m_1$ and $m_2$. $m_2$ is also related with the mass matrix of chargino.
Therefore, the two parameters $m_1$ and $m_2$ can effect the
contributions(neutalino-slepton, chargino-sneutrino) for $\mu\rightarrow e+\gamma$ to some extent.
Supposing $\mu_H=480{\rm GeV}, \tan\beta=11, s_m=3400{\rm GeV}, AL=-2{\rm TeV}, A^{'}_L=300{\rm GeV}$, we
scan the parameters of $m_1$ versus $m_2$ in Fig.\ref{muem1m2}. It implies that in this condition $m_1$ should be in the region $(430\sim630)$GeV
and the effects of $m_2$ are small.

\begin{figure}[h]
\setlength{\unitlength}{1mm}
\centering
\includegraphics[width=3.3in]{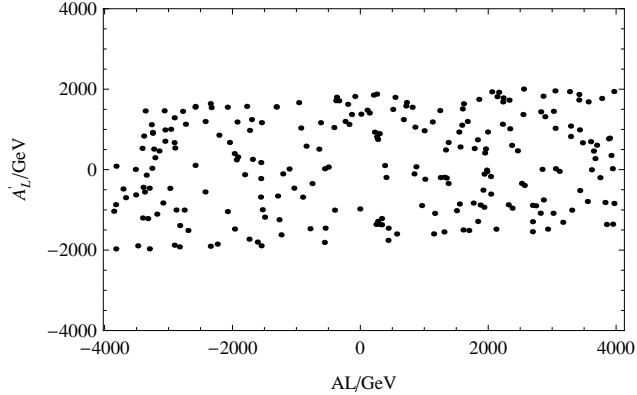}
\caption[]{For $\mu\rightarrow e+\gamma$, the allowed parameters  in the plane of $AL$ versus $A'_L$ with
 $\mu_H=480{\rm GeV}, \tan\beta=12, s_m=3300{\rm GeV}, m_1=500{\rm GeV}, m_2=1{\rm TeV}$.}\label{muealalp}
\end{figure}

The slepton mass squared matrix has $A'_L$ and $AL$ as non-diagonal elements
 which affect the results through slepton-neutralino and slepton-lepton neutralino contributions.
With $\mu_H=480{\rm GeV}, \tan\beta=12, s_m=3300{\rm GeV}, m_1=500{\rm GeV}, m_2=1{\rm TeV}$, in Fig.\ref{muealalp} $A'_L$ versus $AL$ are scanned.
Fig.\ref{muealalp} shows that the effects from $AL$ are smaller than the effects from $A'_L$. The allowed region of $A'_L$ is about $(-2\sim2)$TeV.

\begin{figure}[h]
\setlength{\unitlength}{1mm}
\centering
\includegraphics[width=3.3in]{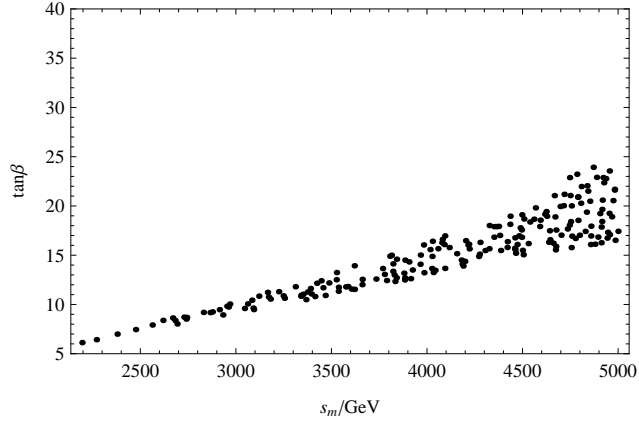}
\caption[]{For $\mu\rightarrow e+\gamma$, the allowed parameters  in the plane of $\tan\beta$ versus $s_m$ with
 $\mu_H=470{\rm GeV}, m_1=500{\rm GeV}, m_2=1{\rm TeV}, AL=-2{\rm TeV}, A^{'}_L=300{\rm GeV}$.}\label{muetbml}
\end{figure}

$\tan\beta$ is related to the mass matrices of chargino, neutralino, slepton and sneutrino, especially to the non-diagonal
elements of these matrices. In BLMSSM, almost all contributions to CLFV processes are influenced by $\tan\beta$. It is a
sensitive parameter and affects the numerical results forcefully.  $m_{\tilde{L}}^2$ and $m_{\tilde{R}}^2$ are introduced
in the soft breaking terms. Both slepton and sneutrino mass squared matrices include $m_{\tilde{L}}^2$ and $m_{\tilde{R}}^2$
, which should give considerable effects to CLFV processes.
Supposing $\mu_H=470{\rm GeV}, m_1=500{\rm GeV}, m_2=1{\rm TeV}, AL=-2{\rm TeV}, A^{'}_L=300{\rm GeV}$, we plot
the results with the allowed $\tan\beta$ versus $s_m$ in Fig.\ref{muetbml}. As we expected, they both are sensitive parameters.
Because the upper bound of $Br(\mu\rightarrow e+\gamma)$ is small, it is easy to exceed the bound
in BLMSSM with the new contributions.

2. $\tau\rightarrow e+\gamma$

\begin{figure}[h]
\setlength{\unitlength}{1mm}
\centering
\includegraphics[width=3.3in]{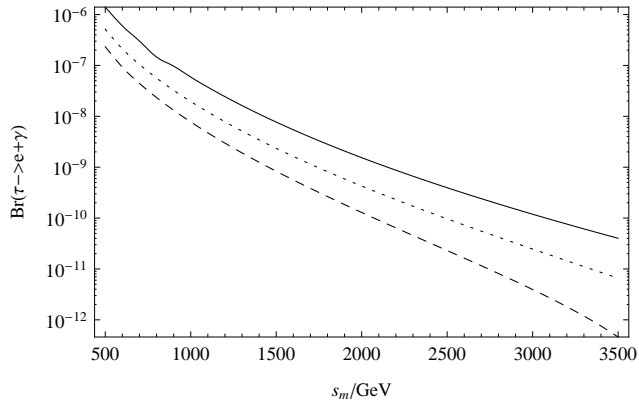}
\caption[]{With $\tan\beta=10(15,25)$, the results of $Br(\tau\rightarrow e+\gamma)$ versus $s_m$
are plotted by the dashed line, dotted line and solid line respectively.}\label{BrTEGMLS}
\end{figure}

The experiment upper bound of $Br(\tau\rightarrow e+\gamma)$ is $3.3\times10^{-8}$ which is almost five-order larger than that of
 $Br(\mu\rightarrow e+\gamma)$. Here we use the parameters $m_2=1{\rm TeV}, \tan\beta_L=2,  M_{sn}=2{\rm TeV}, AL=-2{\rm TeV}, A^{'}_L=300{\rm GeV}, \mu_L=1{\rm TeV},  g_L=\frac{1}{6}$ and $v_{L_t}=3{\rm TeV}$. As discussed in the previous part, $s_m$ can affect the contributions strong through slepton masses. Both
slepton-neutralino and slepton-lepton neutralino give one loop corrections to the CLFV processes.
Using the parameters $m_1=600{\rm GeV},\mu_H=700{\rm GeV}$ and $\tan\beta=10(15,25)$£¬
in Fig.(\ref{BrTEGMLS}) we plot the results varying with $s_m$ by the dashed line, dotted line and solid line.  These three lines are
all decreasing functions of the enlarging $s_m$. In the $s_m$ region $(1000\sim1500){\rm GeV}$, the dashed line varies from $1.0\times10^{-8}$
to $1.0\times10^{-9}$; the solid line varies from $1.0\times10^{-7}$ to $1.0\times10^{-8}$. As $s_m>2500{\rm GeV}$,
the three lines are all much smaller than the upper bound. Corresponding to same $s_m$ during the region $(1\sim2){\rm TeV}$,
the solid line results are about 10 times
as the dashed line results, and the dotted line results are about 3 times as the dashed line results.
It implies that both $s_m$ and $\tan\beta$ are sensitive parameters to the numerical results. Larger $\tan\beta$ leads to
larger results obviously.

\begin{figure}[h]
\setlength{\unitlength}{1mm}
\centering
\includegraphics[width=3.3in]{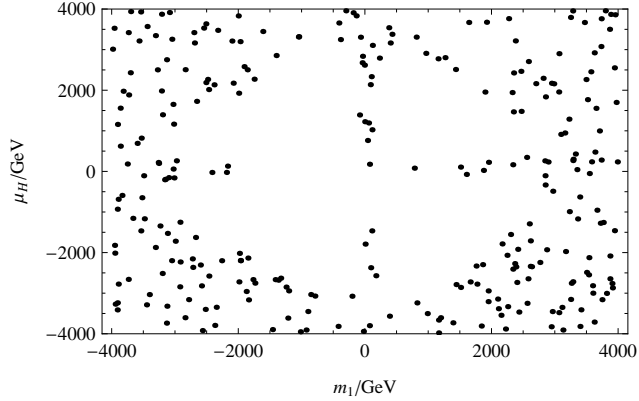}
\caption[]{For $\tau\rightarrow e+\gamma$, the allowed parameters in the plane of $\mu_H$ versus $m_1$ with
 $s_m=800{\rm GeV}, \tan\beta=25$.}\label{tauemum1}
\end{figure}

$\mu_H$ is included in the mass matrices of chargino and neutralino, which should produce considerable influence on the numerical results.
With small $\tan\beta$ and large $s_m$, the results for $\tau\rightarrow e+\gamma$ are much smaller than
the experiment upper bound. To embody effects from $\mu_H$ and $m_1$, we use large $\tan\beta=25$ and small $s_m=800{\rm GeV}$.
The allowed numerical results are plotted by the dots in the Fig.\ref{tauemum1}.

\begin{figure}[h]
\setlength{\unitlength}{1mm}
\centering
\includegraphics[width=3.3in]{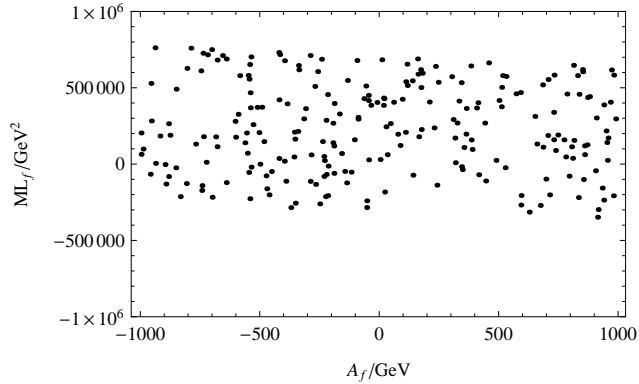}
\caption[]{For $\tau\rightarrow e+\gamma$, the allowed parameters in the plane of $ML_f$ versus $A_f$ with
 $\mu_H=480{\rm GeV}, m_1=800{\rm GeV}, s_m=1500{\rm GeV}, \tan\beta=15$.}\label{tauefmlfal}
\end{figure}

The non-diagonal elements of $(m_{\tilde{L}}^2)$ and $(m_{\tilde{R}}^2)$ represent the lepton flavor mixing and lead to
strong mixing for sleptons (sneutrinos).
To simplify the discussion, the assumption $(m_{\tilde{L}}^2)_{ij}=(m_{\tilde{R}}^2)_{ij}=ML_f$
for $i\neq j$ and $i,j=1,2,3$ is used.
We also consider the non-diagonal elements of $A_l$
with the supposition $(A_l)_{ij}=A_f$ for $i\neq j$ and $i,j=1,2,3$. Using the parameters
$\mu_H=480{\rm GeV}, m_1=800{\rm GeV}, s_m=1500{\rm GeV}, \tan\beta=15$, in the plane of $ML_f$ versus $A_f$
the parameter space is scanned, and the allowed results are shown in Fig.\ref{tauefmlfal}.
The effects from $ML_f$ are stronger than the effects from $A_f$.  Here, the allowed region for $ML_f$ is about
$(-4\times10^5\sim 8\times 10^5)$ GeV$^2$.

3. $\tau\rightarrow \mu+\gamma$

\begin{figure}[h]
\setlength{\unitlength}{1mm}
\centering
\includegraphics[width=3.3in]{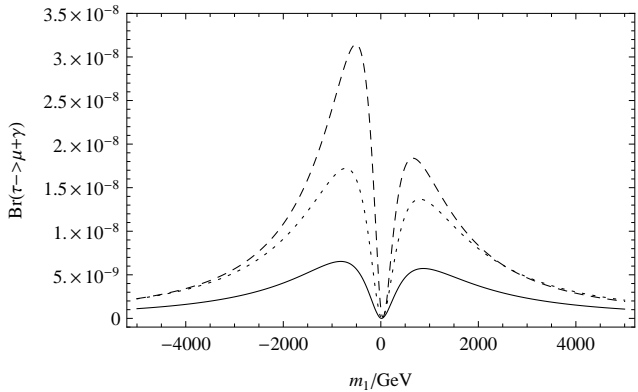}
\caption[]{{With $\mu_H=500(1500,3000){\rm GeV}$, the results of $Br(\tau\rightarrow \mu+\gamma)$ versus $m_1$
are plotted by the dashed line, dotted line and solid line respectively.}}\label{BrTMGM1}
\end{figure}

Similar as $\tau\rightarrow e+\gamma$, the branching ratio of $\tau\rightarrow \mu+\gamma$ is also large, whose
experiment upper bound is $4.4\times10^{-8}$. For the decay $\tau\rightarrow \mu+\gamma$, the parameters
$m_2=1{\rm TeV}, \mu_H=500{\rm GeV},m_1=800{\rm GeV},  M_{sn}=2{\rm TeV}, AL=-2{\rm TeV},
s_m=1{\rm TeV}, A^{'}_L=300{\rm GeV}$ are used.
 The gaugino mass $m_1$ is related to the neutralino-slepton contributions to
CLFV processes. With $\tan\beta=15$, $\mu_H=500(1500,3000){\rm GeV}$, in Fig.(\ref{BrTMGM1}) the results of $Br(\tau\rightarrow \mu+\gamma)$
are studied versus $m_1$ by the
dashed line, dotted line and solid line. As $|m_1|$ is around $600{\rm GeV}$, these three lines arrive at their big values. The results
are decreasing functions of the increasing $|m_1|$, when $|m_1|$ is larger than 800 {\rm GeV}.
The biggest value of the dashed line can reach
$3.2\times 10^{-8}$. The solid line varies from $1.0\times10^{-10}$ to $5.1\times 10^{-9}$.
The dashed line  is the highest line and the
solid line is the lowest one. The dotted line is the middle line and at the order of $10^{-8}$.
The three lines show the CLFV processes are suppressed by heavy virtual particles at several TeV order.

\begin{figure}[h]
\setlength{\unitlength}{1mm}
\centering
\includegraphics[width=3.3in]{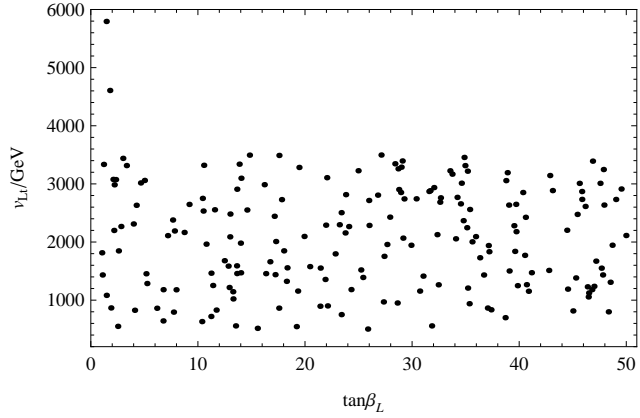}
\caption[]{ For $\tau\rightarrow \mu+\gamma$, the allowed parameters in the plane of $\tan\beta_L$ versus $v_{Lt}$ with
 $g_L=\frac{1}{6}, \mu_L=1{\rm TeV}, \tan\beta=18$.}\label{taumutblvlt}
\end{figure}

Compared with MSSM, $\tan\beta_L$ and $v_{L_t}$ are new parameters that have relation with mass matrices of slepton, sneutrino and lepton neutralino.
Therefore, the effects to CLFV process $\tau\rightarrow \mu+\gamma$ from $\tan\beta_L$ and $v_{L_t}$ are of interest.
Based on the supposition $g_L=\frac{1}{6}, \mu_L=1{\rm TeV}, \tan\beta=18$, we scan the parameter space of $\tan\beta_L$
versus $v_{L_t}$ in Fig.\ref{taumutblvlt}. The value of $\tan\beta_L$ can vary from $(0 \sim 40)$, whose effects are small.
As $\tan\beta_L>2.0$, $v_{L_t}$ should be no more than 3600 GeV.

\begin{figure}[h]
\setlength{\unitlength}{1mm}
\centering
\includegraphics[width=3.3in]{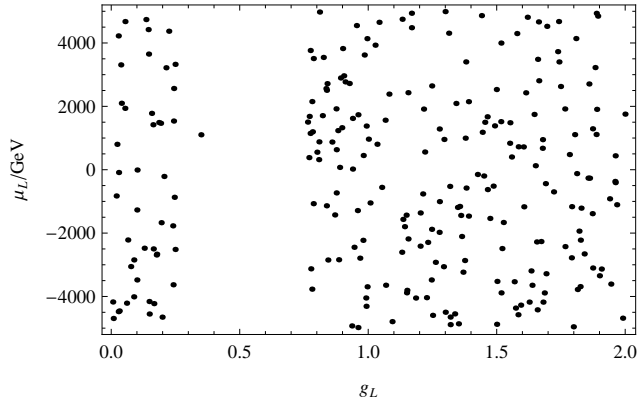}
\caption[]{ For $\tau\rightarrow \mu+\gamma$, the allowed parameters in the plane of $g_L$ versus $\mu_{L}$ with
 $\tan\beta_L=2, v_{L_t}=3{\rm TeV},\tan\beta=18$.}\label{taumumulgl}
\end{figure}

$g_L$ is not only the coupling constant for lepton neutralino-slepton-lepton but also constitute
the mass matrix of lepton neutralino. Considerable influence to $\tau\rightarrow \mu+\gamma$ from $g_L$ is hopeful.
The new gaugino mass $\mu_L$ is the non-diagonal element of the lepton neutralino mass matrix.
To see how $g_L$ and $\mu_L$ affect the numerical results for $\tau\rightarrow \mu+\gamma$,
with $\tan\beta_L=2, v_{L_t}=3{\rm TeV},\tan\beta=18$
we give out the allowed dots in the plane of $g_L$ versus $\mu_L$.
Fig.\ref{taumumulgl} implies that when $g_L$ is near 0.5, the results are larger than the experiment upper bound.
The effects from $\mu_L$ is very weak, and can be neglected.

\subsection{$l_j\rightarrow 3l_i$}

In this subsection, we numerically study the CLFV processes $l_j\rightarrow 3l_i$ with the supposed parameters $m_L=3{\rm TeV},  \mu_L=1{\rm TeV},v_{L_t}=3{\rm TeV}$. These processes have close relations to $l_j\rightarrow l_i+\gamma$.

1. $\mu\rightarrow 3e$

\begin{figure}[h]
\setlength{\unitlength}{1mm}
\centering
\includegraphics[width=3.3in]{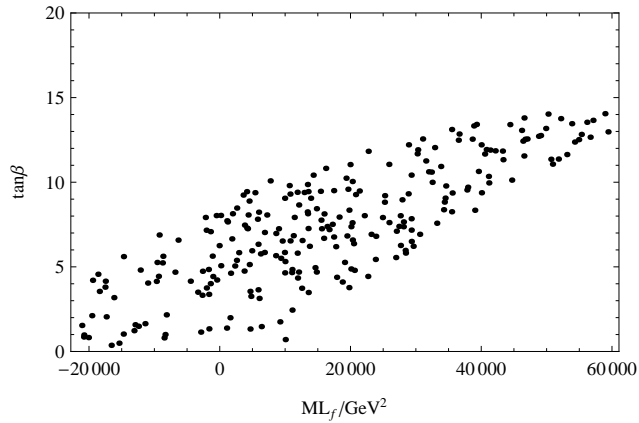}
\caption[]{ For $\mu\rightarrow 3e$, the allowed parameters in the plane of $\tan\beta$ versus $ML_{f}$ with
 $\mu_H=500{\rm GeV}, m_2=1500{\rm GeV}.$}\label{mu3etbmlf}
\end{figure}

The most strict branching ratio of CLFV processes $l_j\rightarrow 3l_i$ is $Br(\mu\rightarrow 3e)$, whose experiment upper bound is
$1.0 \times 10^{-12}$. This experiment constraint is the first one to be considered for $l_j\rightarrow 3l_i$.
To study the process $\mu\rightarrow 3e$, the used
parameters are $M_{sn}=1{\rm TeV}, s_m=3300{\rm GeV},  AN=-500{\rm GeV},  AL=-2{\rm TeV}, A^{'}_L=300{\rm GeV}, m_1=500{\rm GeV},
g_L=1/6,\tan\beta_L=2$. From the discussion in the front section, $\tan\beta$ and non-diagonal elements of $(m_{\tilde{L}}^2)$ and $(m_{\tilde{R}}^2)$
are sensitive parameters to the CLFV processes. With $\mu_H=500{\rm GeV}$ and $m_2=1500{\rm GeV}$, the
parameters $\tan\beta$ versus $ML_f$ are scanned in the Fig.\ref{mu3etbmlf}. The plotted dots represent the allowed results which
embody the  influences from $\tan\beta$ and $ML_f$. Here, the value of $\tan\beta$ should be no more than 15.

\begin{figure}[h]
\setlength{\unitlength}{1mm}
\centering
\includegraphics[width=3.3in]{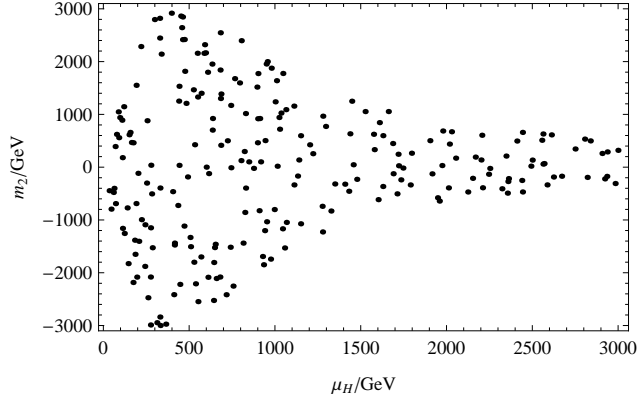}
\caption[]{  For $\mu\rightarrow 3e$, the allowed parameters in the plane of $m_2$ versus $\mu_H$ with
 $ML_f=10^4{\rm GeV^2},\tan\beta=10$.}\label{mu3em2mu}
\end{figure}

Here we consider the non-diagonal elements of $(m_{\tilde{L}}^2)$ and $(m_{\tilde{R}}^2)$, and suppose $ML_f=10^4{\rm GeV^2}$ and $\tan\beta=10$.
After the numerical calculation, the allowed parameters in the plane of $m_2$ versus $\mu_H$ are shown in the Fig.\ref{mu3em2mu}.
When $\mu_H$ is near 500 GeV, $m_2$ can vary from -3 TeV to 3 TeV. As
$\mu_H>600$ GeV, the allowed scope of $m_2$ shrinks with the enlarging $\mu_H$.

2. $\tau\rightarrow 3e$

\begin{figure}[h]
\setlength{\unitlength}{1mm}
\centering
\includegraphics[width=3.3in]{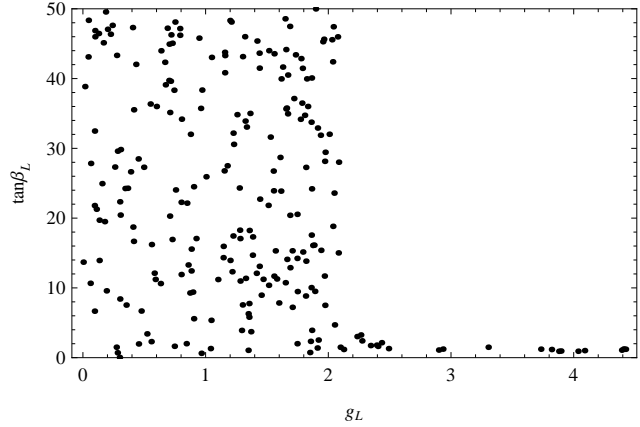}
\caption[]{  For $\tau\rightarrow 3e$, the allowed parameters in the plane of $\tan\beta_L$ versus $g_L$ with
 $s_m=3500{\rm GeV}, AN=-500{\rm GeV},  AL=-2{\rm TeV}, A^{'}_L=300{\rm GeV}$.}\label{tau3egltbl}
\end{figure}

The experiment upper bound of the CLFV process $Br(\tau\rightarrow 3e)$ is $(2.7\times10^{-8})$, and it
is about four order larger than that of $\mu\rightarrow 3e$.
 For the study of
$\tau\rightarrow 3e$, $\tan\beta=10, M_{sn}=2{\rm TeV}, \mu_H=3{\rm TeV},m_2=1500{\rm GeV},  m_1=2500{\rm GeV}$ are supposed here.
To show the importance
of the non-MSSM contributions from lepton neutralino-selepton, we discuss the effects from $g_L$ and $\tan\beta_L$
 with $s_m=3500{\rm GeV}, AN=-500{\rm GeV},  AL=-2{\rm TeV}, A^{'}_L=300{\rm GeV}$.
Fig.\ref{tau3egltbl} implies in the $g_L$ region $(0\sim 2)$, the allowed scope of $\tan\beta_L$ is from $0$ to 50.
When $g_L$ is larger than 2.2, the region of $\tan\beta_L$ turns very small which is just from 0 to 2.

\begin{figure}[h]
\setlength{\unitlength}{1mm}
\centering
\includegraphics[width=3.3in]{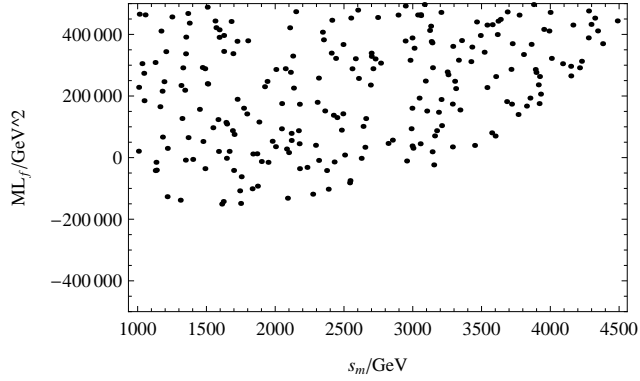}
\caption[]{For $\tau\rightarrow 3e$, the allowed parameters in the plane of $s_m$ versus $ML_f$ with
 $AN=2{\rm TeV},  AL=2{\rm TeV}, A^{'}_L=500{\rm GeV},g_L=0.1,\tan\beta_L=1$.}\label{tau3emldfd}
\end{figure}

$(m_{\tilde{L}}^2)$ and $(m_{\tilde{R}}^2)$ are sensitive parameters relating to
lepton mixing between different generations. That is to say, their diagonal and non-diagonal elements are all
important factors for CLFV processes.
 As $AN=2{\rm TeV},  AL=2{\rm TeV}, A^{'}_L=500{\rm GeV},g_L=0.1,\tan\beta_L=1$, the allowed
 scope of $s_m$ versus $ML_f$ is plotted in the Fig.\ref{tau3emldfd}.
  $ML_f$ should be no less than $-2.0\times10^{5} {\rm GeV^2}$ and
    the allowed
  smallest values of $ML_f$ turn large with the enlarging $s_m$.

3. $\tau\rightarrow 3\mu$

\begin{figure}[h]
\setlength{\unitlength}{1mm}
\centering
\includegraphics[width=3.3in]{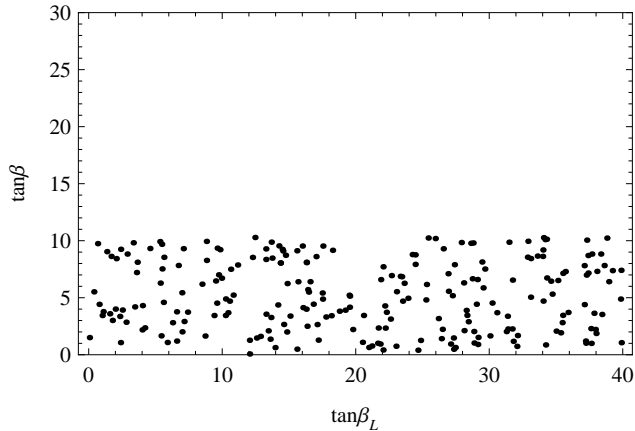}
\caption[]{For $\tau\rightarrow 3\mu$, the allowed parameters in the plane of $\tan\beta$ versus $\tan\beta_L$ with
 $s_m=1500{\rm GeV}$.}\label{tau3mutbtbl}
\end{figure}

Similarly, we calculate the CLFV process $\tau\rightarrow 3\mu$, whose experiment upper bound is $Br(\tau\rightarrow 3\mu)<2.1\times10^{-8}$.
To obtain the numerical results for $\tau\rightarrow 3\mu$, we use
$M_{sn}=1{\rm TeV}, \mu_H=500{\rm GeV},m_2=1500{\rm GeV},  m_1=2500{\rm GeV},AN=3{\rm TeV},  AL=3{\rm TeV}, A^{'}_L=3{\rm TeV},
 g_L=0.4$. In the studied processes, there are two angles $\tan\beta$ and $\tan\beta_L$ relating to the contributions.
In the plane of $\tan\beta$ versus $\tan\beta_L$, with $s_m=1500{\rm GeV}$  we show the allowed results denoted by the dots in Fig.\ref{tau3mutbtbl}.
The suitable value of $\tan\beta$ is in the region $(0\sim 10)$. Compared with the effects from $\tan\beta$,
those effects from $\tan\beta_L$ are tiny.

\begin{figure}[h]
\setlength{\unitlength}{1mm}
\centering
\includegraphics[width=3.3in]{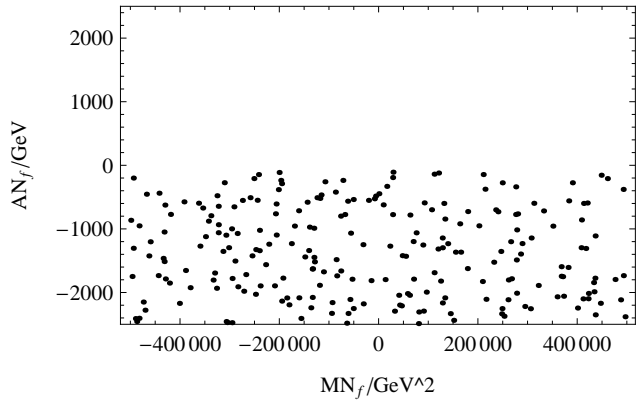}
\caption[]{For $\tau\rightarrow 3\mu$, the allowed parameters in the plane of $AN_f$ versus $MN_f$ with
 $\tan\beta=10, \tan\beta_L=10, s_m=2{\rm TeV}$.}\label{tau3muanfmnf}
\end{figure}

The mass squared  matrix of sneutrino include $A_N$, $A_{N^c}$ and $m^2_{\tilde{N}^c}$,
which naturally influence the contributions from sneutrinos and charginos.
We take into account the non-diagonal elements of these parameters and suppose
 $(A_N)_{ij}=(A_{N^c})_{ij}=AN_f,\;(m^2_{\tilde{N}^c})_{ij}=MN_f$ for $i\neq j$ and $i,j=1,2,3$.
With $\tan\beta=10, \tan\beta_L=10, s_m=2{\rm TeV}$, in Fig.\ref{tau3muanfmnf} we plot the allowed results in the plane of $AN_f$ versus $MN_f$.
To obtain the allowed results, $AN_f$ is no more than zero. The effects from $MN_f$ are tiny and ignorable.

\section{discussion and conclusion}
In SM the theoretical predictions for CLFV processes $l_j\rightarrow l_i+\gamma$ and $l_j\rightarrow 3l_i$
are much smaller than their experiment upper bounds. If large branching ratios of CLFV processes are detected,
there must be new physics beyond SM. In BLMSSM, there are new parameters and new contributions to CLFV processes.
For example, beside three light neutrinos there are three heavy neutrinos and six sneutrinos.
Furthermore, new gauginos and new higgsinos mix leading to three lepton neutralinos, that can give new type contributions
through the coupling of lepton neutralino-slepton-lepton.

The branching ratio experiment upper bounds of $\mu\rightarrow e+\gamma$ and $\mu\rightarrow 3e$ are strict and rigorously confine the parameter space.
For the both processes, it is very easy to exceed their experiment upper bounds. The experiment upper bounds of
the processes $\tau\rightarrow e+\gamma, \tau\rightarrow \mu+\gamma,
\tau\rightarrow 3e$ and $\tau\rightarrow 3\mu$  are all at the order of $10^{-8}$ and much larger than
those of $\mu\rightarrow e+\gamma$ and $\mu\rightarrow 3e$. In our used parameter space of BLMSSM, the branching ratios of these six CLFV processes can be large enough
to achieve the bounds and even surpass them.  From the numerical results, one finds the
important parameters are $\tan\beta, s_m,ML_f,m_1,m_2,\mu_H$ and $g_L$, where $\tan\beta, s_m, ML_f$ are very sensitive parameters.
We hope in the near future large branching ratios of CLFV processes can be detected by the experiments.

{\bf Acknowledgements}

   This work has been supported by the Major Project of NNSFC(NO.11535002) and NNSFC(NO.11275036),
   the Open Project Program of State Key Laboratory of Theoretical Physics, Institute of Theoretical Physics, Chinese
Academy of Sciences, China (No.Y5KF131CJ1), the Natural Science Foundation
of Hebei province with Grant No. A2013201277, and the Found of Hebei province with
the Grant NO. BR2-201 and the
Natural Science Fund of Hebei University with Grants
No. 2011JQ05 and No. 2012-242, Hebei Key Lab of Optic-Electronic Information and Materials,
 the midwest universities comprehensive strength promotion project.


\begin{thebibliography}{99}

\bibitem{neutrino} K. Abe et al.(T2K Collaboration), Phys. Rev. Lett. {\bf107}
(2011) 041801; J. Ahn et al. (RENO Collaboration), Phys. Rev. Lett. {\bf108} (2012) 191802.
, F.An et al. (DAYA-BAY Collaboration),
Phys. Rev. Lett. {\bf108} (2012) 171803.

\bibitem{neutrinoN1}E. Ma, A. Natale, O. Popov, Phys. Lett. B{\bf746} (2015) 114-116;
I. Girardi , S.T. Petcov , A.V. Titov, Nucl. Phys. B {\bf894} (2015) 733-768.
\bibitem{neutrinoN2}
P. Ghosh, S. Roy, JHEP {\bf0904} (2009) 069;
P. Ghosh, P. Dey, B. Mukhopadhyaya, S. Roy, JHEP {\bf1005} (2010) 087.

\bibitem{mueg}S. Petcov, Sov.J.Nucl.Phys. {\bf 25} (1977) 340.

\bibitem{14pdg} K.A. Olive et al. (Particle Data Group), Chin. Phys. C {\bf 38} (2014) 090001.

\bibitem{RNeutrino}T. Goto, Y. Okada, T. Shindou, M. Tanaka, R. Watanabe,
 Phys. Rev. D {\bf 91} (2015) 033007.

\bibitem{MSSM} J. Rosiek, Phys. Rev. D {\bf 41} (1990) 3464 [Erratum hep-ph/9511250].

\bibitem{nuRMSSM}A. Ilakovac, A. Pilaftsis, L. Popov, Phys. Rev. D{\bf 87} (2013) 053014.

\bibitem{ljlig}J. Hisano, T. Moroi, K. Tobe, M. Yamaguchi, Phys. Rev. D {\bf53} (1996) 2442.
\bibitem{ZHB}Hai-Bin Zhang, Tai-Fu Feng, Li-Na Kou, Shu-Min Zhao, Int.J.Mod.Phys. A {\bf28} (2013) 24, 1350117;
Hai-Bin Zhang, Tai-Fu Feng, Shu-Min Zhao, Tie-Jun Gao,  Nucl.Phys. B {\bf873} (2013) 300-324, Errutum: Nucl. Phys. B {\bf879} (2014) 235.


\bibitem{Rp} H.P. Nilles, Phys. Rep. {\bf 110} (1984) 1.
 \bibitem{BLMSSM}P. F. Perez, Phys. Lett. B {\bf711} (2012) 353; J. M. Arnold, P. F. Perez, B. Fornal,
 and S. Spinner, Phys. Rev. D {\bf 85} (2012)115024.
 \bibitem{Rp1} R. Barbieri,  A. Masiero, Nucl. Phys. B {\bf 267} (1986) 679;
S. Dimopoulos, L.J. Hall, Phys. Lett. B {\bf207} (1987) 210.
\bibitem{BLMSSM1} P. F. Perez and M. B. Wise, JHEP {\bf 1108} (2011) 068; Phys. Rev. D {\bf 82} (2010) 011901.


\bibitem{weBLMSSM}Tai-Fu Feng, Shu-Min Zhao, Hai-Bin Zhang, et al., Nucl. Phys. B {\bf 871} (2013) 223.
\bibitem{smneutron}Shu-Min Zhao, Tai-Fu Feng, Ben Yan et al., JHEP {\bf 1310} (2013) 020.
\bibitem{BLLEDM}Shu-Min Zhao, Tai-Fu Feng, Xi-Jie Zhan, Hai-Bin Zhang, Ben Yan, JHEP {\bf 1507} (2015) 124.
\bibitem{sunfei}Fei Sun, Tai-Fu Feng, Shu-Min Zhao et al., Nucl. Phys. B {\bf 888} (2014) 30; .
Tie-Jun Gao, Tai-Fu Feng, Fei Sun, Commun. Theor. Phys. {\bf61} (2014) 1, 95-100;
Fei Sun, Tai-Fu Feng , Tie-Jun Gao, Hai-Bin Zhang, Shu-Min Zhao, Int.J.Mod.Phys. A {\bf29} (2014) 27, 1450153.
\bibitem{Higgs}CMS Collaboration, Phys. Lett. B {\bf 716} (2012) 30; ATLAS Collaboration, Phys. Lett. B {\bf 716} (2012) 1;
CMS Collaboration arXiv:hep-ph/1301.3405.

\bibitem{BLMass}
J.M. Arnold, P.F. Perez, B. Fornal, S. Spinner, Phys. Rev. D {\bf85} (2012) 115024.

\bibitem{BiaoChen} Chen Biao, Zhao Shu-min, Yan Ben, et al., Commun. Theor. Phys. {\bf61} (2014) 619-623;
Shu-Min Zhao, Tai-Fu Feng, Hai-Bin Zhang et al., JHEP  {\bf1411} (2014) 119.


\end{thebibliography}
\end{document}